\documentclass[prd,tightenlines,nofootinbib,showpacs,preprintnumbers,superscriptaddress, twocolumn]{revtex4}
\usepackage{amsfonts,amsmath,amssymb,amsthm,bbm,hyperref}
\usepackage{graphicx}
\usepackage{color}
\usepackage[T1]{fontenc}

\newcommand{\R}{{\mathbb R}}

\newcommand{\be}{\begin{equation}}
\newcommand{\ee}{\end{equation}}
\newcommand{\beq}{\begin{eqnarray}}
\newcommand{\eeq}{\end{eqnarray}}
\newcommand{\nn}{\nonumber}

\DeclareMathOperator\arctanh{arctanh}
\DeclareMathOperator\cotanh{cotanh}

\newcommand{\vphi}{\varphi}

\begin{document}

\title{Quantum cosmology of a conformal multiverse}
\author{Salvador J. Robles-P\'{e}rez}
\affiliation{Instituto de F\'{\i}sica Fundamental, Consejo Superior de Investigaciones Cient\'{\i}ficas, Serrano 121, 28006 Madrid, Spain,}
\affiliation{Estaci\'{o}n Ecol\'{o}gica de Biocosmolog\'{\i}a, Pedro de Alvarado, 14, 06411-Medell\'{\i}n, Spain.}
\date{\today}

\begin{abstract}
In this paper it is studied the cosmology of a homogeneous and isotropic spacetime endorsed with a conformally coupled massless scalar field. We find six different solutions of the Friedmann equation that represent six different types of universes, all of them are periodically distributed along the complex time axis. From a classical point of view, they are then isolated, separated by Euclidean regions that represent quantum mechanical barriers. Quantum mechanically, however, there is a non-zero probability for the state of the universes to tunnel out through a Euclidean instanton and suffer a sudden transition to another state of the spacetime. We compute the probability of transition for this and other non-local processes like the creation of universes in entangled pairs and, generally speaking, in multipartite entangled states. We obtain the quantum state of a single universe within the formalism of the Wheeler-DeWitt equation and give the semiclassical state of the universes that  describes the quantum mechanics of a scalar field propagating in a deSitter background spacetime. We show that the superposition principle of the quantum mechanics of matter fields alone is an emergent feature of the semiclassical description of the universe that is not valid, for instance, in the spacetime foam. 

We use the third quantization formalism to describe the creation of an entangled pair of universes with opposite signs of their momenta conjugated to the scale factor. Each universe of the entangled pair represents an expanding spacetime in terms of the WKB time experienced by internal observers in their particle physics experiments. We compute the effective value of the Friedmann equation of the background spacetime of the two entangled universes and, thus, the effects that the entanglement would have in their expansion rates. We analyze as well the effects of the inter-universal entanglement in the properties of the scalar fields that propagate in each spacetime of the entangled pair. We find that the largest modes of the scalar field are unaware of the entanglement between universes but the effects can be significant for the lowest modes, allowing us to compute, in principle, more detailed observational imprints of the multiverse in the properties of a single universe like ours.
\end{abstract}

\maketitle


\section{Introduction}

Quantum cosmology \cite{Hawking1983, Coleman1990, GellMann1990, Vilenkin1994, Vilenkin1995, Wiltshire2003, Kiefer2007} is the application of the quantum theory to the universe as a whole. As a first approximation, this  can be modeled by a composite system of spacetime and matter fields that, at least in the semiclassical regime, can be described as two weakly coupled subsystems. On the other hand, quantum theory is essentially a non local theory (in the sense of Refs. \cite{Einstein1935, Bell1987}) and thus, quantum cosmology inevitably leads  to the concept of a wave function of the universe \cite{Hartle1983, Kiefer1994}. However, the universe is not the whole thing in modern cosmology. 

The multiverse, in its wide variety of forms \cite{Everett1957, DeWitt1973, Linde1983, Linde1986, Smolin1997, Steinhard2002, Susskind2003, Tegmark2003, Freivogel2004, Tegmark2007, Carr2007, Mersini2008a, Mersini2008b, RP2010, Alonso2012}, has become the most general scenario in cosmology. Three of the major frameworks in cosmology, i.e. the landscape of the string theories \cite{Susskind2003, Freivogel2004}, the inflationary paradigm that leads to a continuing generation of inflating bubbles \cite{ Linde1983, Linde1986}, and quantum cosmology \cite{Everett1957, RP2010}, not only support but even enhance the consideration of a multiverse. Even more, these three types of multiverse are complementary so one can generally consider an interacting multiverse \cite{RP2013, RP2016}, which consists of the landscape populated of inflating bubbles that are quantum mechanically described by wave functions that can interfere and where non-local interactions may generally exist. In that case, the same reasoning based on the non locality of the quantum theory that leads to the concept of the wave function of the universe leads now to the concept of the wave function of the multiverse \cite{RP2010}. 

As it happens in quantum mechanics, where one has to work in the framework of a quantum field theory to better describe a many particle system, the quantum state of the multiverse is better described in the framework of the so called third quantization formalism \cite{Strominger1990, RP2010, RP2013}, where it can be defined quantum operators that describe the creation and the annihilation of universes. It parallels the formalism of a quantum field theory, actually, but the field to be quantized  now is the wave function of the multiverse and the abstract space where it \emph{propagates} is the superspace of geometries and matter fields (to follow a quick analogy see, for instance, the table of p. 295 in Ref. \cite{Strominger1990}).

One of the main aims of quantum cosmology is to describe the creation of the universe \cite{Hawking1982, Vilenkin1982, Hartle1983, Vilenkin1984, Linde1991, Mersini2008a}. It is a key feature because once the creation of the universe is described, the description of the rest of physical processes in the universe follows from known physical laws. For instance,  the Schr\"{o}dinger equation of matter fields can be derived in the semiclassical regime from the Wheeler-DeWitt equation \cite{Hartle1993, GellMann1993, Kiefer1998, Joos2003, Kiefer2007}. Then, the wave function of the universe turns out to describe a set of matter fields propagating, and generally interacting, in a curved background spacetime. In that sense, the wave function of the universe contains all the information about particle physics and, of course, classical mechanics too\footnote{This is not surprising as quantum cosmology is constructed from the action that represents the spacetime of general relativity and matter fields propagating therein. Then, it is not strange that a \emph{top-down} approach can derive general relativity (and classical mechanics) and the physics of matter fields from quantum cosmology.}.

In quantum cosmology the customary picture for the creation of the universe is the creation of the universe from nothing \cite{Hawking1982, Hartle1983, Hawking1983, Vilenkin1982, Vilenkin1984}, where by nothing it is not understood the absolute meaning of nothing, i.e. something to which ascribe no properties, but the Euclidean region of the spacetime where nothing real exists. In particular, time does not exist and  no classical physics can therefore be developed. Hence, the Euclidean region represents a quantum barrier for the classical spacetime. However, we know from quantum mechanics that there is generally a non zero probability for a wave function to penetrate into a quantum barrier. Then, there is a non zero probability for the universe to be created from \emph{nothing}. That is all we need to exist.

Nevertheless, the reasoning that leads Hartle and Hawking \cite{Hartle1983} to propose the creation of the universe from nothing might not be correct. As Gott and Li first pointed out \cite{Gott1998} and Barvinsky and Kamenshchik have shown later on \cite{Barvinsky2006, Barvinsky2007a}, the renormalization of the matter fields does not need to exactly balance their zero point vacuum energy. In that case, there are two ways in which the universes can be created, either they are created from a pre-existing baby universe, which  would lead to an eternal and self-contained multiverse \cite{Gott1998}, or they are created from nothing but then, they have to be created in entangled pairs \cite{RP2014}.

In this paper, we shall explore the observable consequences that the creation of universes in entangled pairs might leave in the  properties of a universe like ours. Not until recent years it was generally thought that even if other universes would exist, it would be meaningless to ask for the imprints that they could leave in the observable properties of our universe because  the definition of the universe always entails some notion of causal closure. However, this notion of causal closure is always defined in a local sense (which is the only sense in which causality is properly defined in physics), in terms of the structure of the light ray cones of a given spacetime and the link between causal events. Non-local correlations may still be present in the multiverse without violating the local notion of causal closure, and they might modify some properties of the universe that would leave observable consequences. Thus, the search of the observable imprints of the multiverse in the properties of our universe has become an intense subject of research \cite{Mersini2008, Maldacena2013, Bousso2014, Bousso2015, Kanno2015, Garriga2016, RP2016, Mersini2017, Divalentino2017a, Divalentino2017b, RP2017b}.

Besides, quantum entanglement \cite{Horodecki2009} has a property that is specially appealing in quantum cosmology and the physics of matter fields in curved spacetimes. It is the relationship with the information that we can obtained from a physical subsystem \cite{Vedral2006, Jaeger2007}. It is sometimes stated that information may disappear under some physical processes but entanglement may be telling us that part of the initial information can fall into regions that are inaccessible to us. In that case, it would be a lost of information but not a \emph{fundamental} lost of information.

For instance, general relativity and quantum cosmology are invariant under general spacetime diffeomorphisms. In particular, the quantum state of the universe \cite{Hartle1983} is invariant under a time reversal change. However, the semiclassical state of the universe, in which it is defined the physics we know, has one definite direction of time \cite{Hawking1985, Halliwell1996, Kiefer1995}. The processes occurring in the opposite direction of time seem to have disappeared in the actual universe. However, entanglement may be telling us that they have not disappeared but they can be in a region of the spacetime that is not accessible for us. In fact, the time reversal invariance of the spacetime is broken in the semiclassical universe but a time symmetric solution always coexists because the time reversal invariance of the Friedmann equation. Therefore, if one consider that these two universes are created in entangled pairs, then, the time reversal symmetry does not disappear, it only lives in an inaccessible region\footnote{We are not considering here the thermodynamical arrow of time but just the cosmological one.}.

In this paper we study, in the context of the quantum multiverse, the cosmology of a homogeneous and isotropic spacetime with a conformally coupled massless scalar field, for which exact analytical solutions can be given. We shall show that the most natural way in which the universes are created is in entangled pairs that would conserve the total momentum conjugated to the scalar field. It parallels the creation of particles in entangled pairs with opposite momenta in a quantum field theory. However, the momentum conjugated to the scale factor depends on the expansion rate of the universe and thus, the universes are created with opposite expansion rates from the point of view of a common time variable. Nevertheless, in terms of the time variable measured in particle experiments made by internal observers the entangled universes are both expanding or contracting. We shall then  study the properties of entanglement between: i) the correlated wave functions of the spacetimes, and ii) the states of the scalar fields that propagate therein.

The outline is the following. In Sect. II, we obtain the classical solutions of the Friedmann equation. They represent six different types of universes in the Lorentzian sector, and  different Euclidean instantons in the Euclidean sector. In Sect. III, we obtain the quantum state of a single universe by solving the Wheeler-DeWitt equation. We compute as well the probabilities of transition between different states of the universe and the probabilities for the universes to be created in entangled pairs and, generally speaking, in multipartite entangled states. In Sect. IV.A we apply the third quantization formalism to describe the creation of universes in entangled pairs and to compute the effect of this inter-universal entanglement in the effective value of the Friedmann equation. In Sect. IV.B we compute the effects that the entanglement between two spacetimes would have in the properties of the scalar fields that propagate in the entangled spacetimes. Finally, in Sect. V we draw some conclusions and make further comments about the observability of the multiverse.

\section{Classical states}

\subsection{Friedmann equation}

The Einstein-Hilbert action for a homogeneous and isotropic spacetime, with metric  \cite{Hartle1983}
\be\label{MT01}
ds^2 = \sigma^2 \left( - N^2(t) dt^2 + a^2(t) d\Omega_3^2 \right) ,
\ee
where $N(t)$ is the lapse function and, $\sigma^2 = l^2/24\pi^2$, and a conformally coupled massless scalar field, $\vphi$, can be written as
\be\label{ACT01}
S = \frac{1}{2} \int dt \ N \left( - \frac{a \dot{a}^2}{N^2} + a - \lambda a^3 + \frac{a \dot{\chi}^2}{N^2} - \frac{1}{a} \chi^2 \right) , 
\ee
where, $\dot{a} = \frac{d a}{d t}$, $\lambda \equiv \frac{\sigma^2 \Lambda}{3} \equiv H^2$, and the field has been rescaled according to \cite{Hartle1983}
\be\label{VPHI01}
\vphi = \frac{\chi}{(2\pi^2\sigma^2)^\frac{1}{2}a} .
\ee
In conformal time, defined by
\be
d\eta = \frac{d t}{a} ,
\ee
the action (\ref{ACT01}) turns out to be
\be
S = \frac{1}{2} \int d\eta \ N \left( - \frac{(a')^2}{N^2} + a^2 - H^2 a^4 + \frac{(\chi')^2}{N^2} -  \chi^2 \right) , 
\ee
where the prime denotes derivative with respect to the conformal time, i.e. $a' = \frac{da}{d\eta}$ and $\chi' = \frac{d\chi}{d\eta}$. The invariance of the action under reparametrizations of time gives rise to the Hamiltonian constraint, which reads
\be\label{HAM01}
H = \frac{N}{2 a} \left( - p_a^2 - a^2 + H^2a^4 + p_\chi^2 + \chi^2 \right) = 0 ,
\ee
with, 
\be\label{M01}
p_a \equiv \frac{\delta L}{\delta\dot{a}} = -\frac{a \dot{a}}{N} \ , \ p_\chi \equiv\frac{\delta L}{\delta \dot{\chi}} = \frac{a \dot{\chi}}{N} ,
\ee
or
\be\label{HAM02}
H = \frac{N}{2} \left( - \pi_a^2 - a^2 + H^2a^4 + \pi_\chi^2 + \chi^2 \right) = 0 ,
\ee
with, 
\be\label{M02}
\pi_a \equiv \frac{\delta L}{\delta a'}= -\frac{ a' }{N}  \ , \ \pi_\chi \equiv\frac{\delta L}{\delta \chi'} = \frac{ \chi'}{N}.
\ee
Eqs. (\ref{HAM01}) and (\ref{HAM02}) reflect the invariance of the Hamiltonian constraint under time reparametrizations too. In fact, we can always consider the canonical transformation given by
\be
(a,\pi_a,\chi,\pi_\chi; \eta) \rightarrow (\tilde{a}= a, \tilde{\pi}_a = \pi_a, \tilde{\chi} =\chi, \tilde{\pi}_\chi = \pi_\chi; \xi) ,
\ee
with
\be
\xi = \int N d\eta 
\ee
so that
\be
H_\xi(\tilde{a}, \tilde{\pi}_a, \tilde{\chi}, \tilde{\pi}_\chi; \xi) \frac{d\xi}{d\eta} = H(a,\pi_a,\chi,\pi_\chi;\eta)  ,
\ee
and
\be\label{HAM03}
H_\xi = \frac{1}{2} \left( - \pi_a^2 - a^2 + H^2a^4 + \pi_\chi^2 + \chi^2 \right) .
\ee
In this paper, unless otherwise indicated, we shall assume that $N=1$, for which $\xi = \eta$ is the conformal time, and $t$ is cosmic time. The Hamiltonian constraint shows that the total energy of the universe is zero,
\be\label{HAMcons}
H_T = - H_a + H_\chi = 0,
\ee
with
\beq\label{HAMchi}
H_\chi &=& \frac{1}{2} \pi_\chi^2 + \frac{1}{2} \chi^2 , \\ \label{HAMa}
H_a &=& \frac{1}{2} \pi_a^2 + \frac{1}{2} a^2 - \frac{H^2}{ 2} a^4 .
\eeq
The first of these Hamiltonians corresponds to the Hamiltonian of a harmonic oscillator with unit frequency and equation of motion
\be\label{WE01}
\chi''(\eta) + \chi(\eta) = 0 ,
\ee
which can be integrated (after multiplying by $\chi'(\eta)$) to yield
\be
\frac{1}{2} \chi'^2 + \frac{1}{2} \chi^2 = E ,
\ee
where $E$ is the conserved energy of the scalar field. The Hamiltonian (\ref{HAMa}) is the Hamiltonian of an anharmonic oscillator with equation of motion
\be
a''(\eta) + a(\eta) - 2 H^2 a^3(\eta) = 0 ,
\ee
that can be integrated as well by multiplying it by $a'$ to yield
\be\label{EQMa}
\frac{1}{2} a'^2 + \frac{1}{2} \left(a^2 - H^2 a^4 \right) = \varepsilon ,
\ee
where $\varepsilon$ is a constant too. In order to satisfy the Hamiltonian constraint (\ref{HAMcons}),
\be
H_T = - \varepsilon + E = 0 .
\ee
It turns out that, $\varepsilon = E$, i.e. the energy of the spacetime is negative and equals the energy of the scalar field so the total energy of the universe is zero. Eq. (\ref{EQMa}) can also be expressed as
\be\label{FE00}
a'^2 = H^2a^4 - a^2 + 2 E ,
\ee
which is nothing more than the Friedmann equation written in conformal time. In cosmic time it turns out to be
\be\label{FE01}
\dot{a}^2 = H^2 a^2 - 1 + \frac{C}{a^2}   ,
\ee
where \cite{Gott1998}, $C \equiv 2 E = \frac{8 \pi \rho a^4}{3}$, with $\rho$ being the energy density of the conformally coupled scalar field.

\begin{figure}
\centering
\includegraphics[width=8cm]{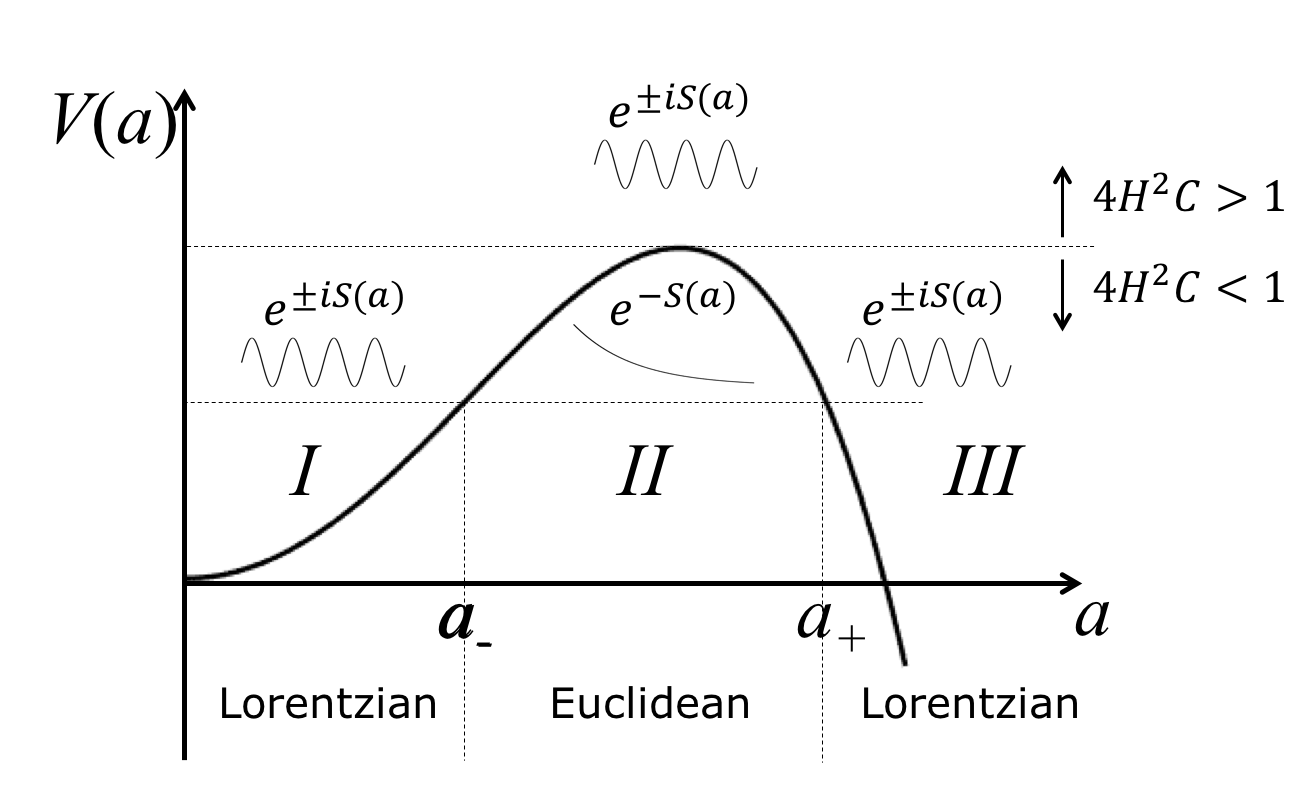}
\caption{Lorentzian and Euclidean regions for the potential $V(a) = a^2 - H^2 a^4$ in (\ref{EQMa}), with $C = 2 \varepsilon$. For the value, $4H^2 C > 1$, there is no forbidden region. For the value, $4 H^2 C < 1$, there are two classically allowed regions separated by a quantum barrier.}
\label{figure01}
\end{figure}

\subsection{Lorentzian solutions}

Let us now find the Lorentzian solutions of the Friedmann equation (\ref{FE01}), which can also be written as
\be\label{FE02}
\frac{da}{dt} = \frac{H}{a} \sqrt{(a^2-a_+^2)(a^2 - a_-^2)} ,
\ee
with \cite{Gott1998}
\be\label{apm}
a_\pm^2 = \frac{1}{2H^2} \left( 1\pm \left(1 - 4 C H^2 \right)^\frac{1}{2} \right)  .
\ee
For the value, $4 H^2 C \geq 1$,  the r.h.s of (\ref{FE01}) is always non negative and there is thus a Lorentzian solution of the Friedmann equation for all values of the scale factor. For $4 H^2 C < 1$, however, it is only positive for the values, $a > a_+$ and $a<a_-$. These two regions represent classically allowed regions for which an analytical Lorentzian solution can be given. They are separated by the Euclidean region, $a_- < a < a_+$, for which exists no real solution of (\ref{FE01}). We have therefore two types of universes separated by a quantum barrier (see Fig. \ref{figure01}).

\begin{figure}
\centering
\includegraphics[width=8cm]{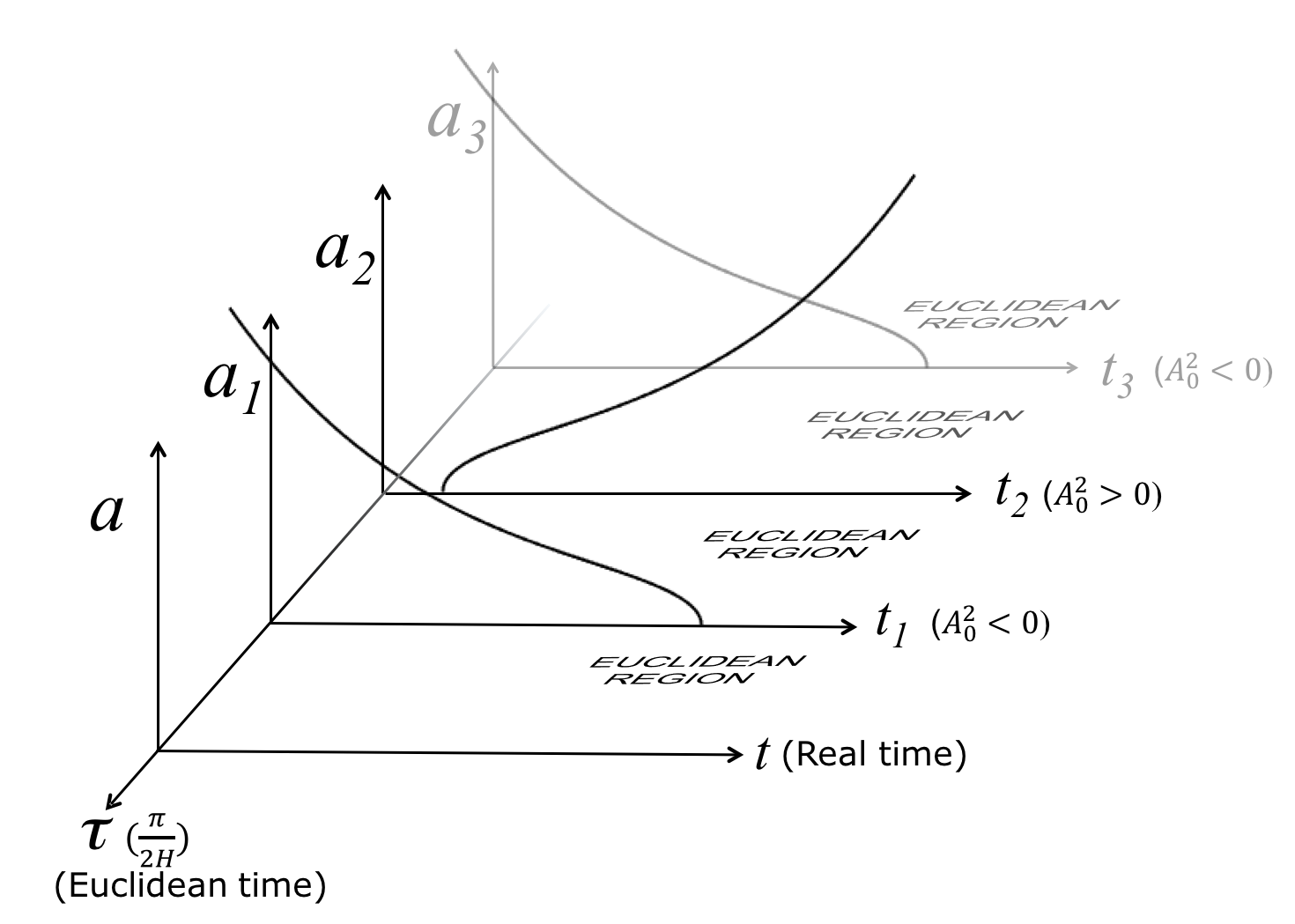}
\caption{Solution (\ref{SF11}) of the Friedmann equation for $A_0^2 = \pm 1$ and, $4 H^2 C > 1$.}
\label{figure02}
\end{figure}

\begin{figure}
\centering
\includegraphics[width=8cm]{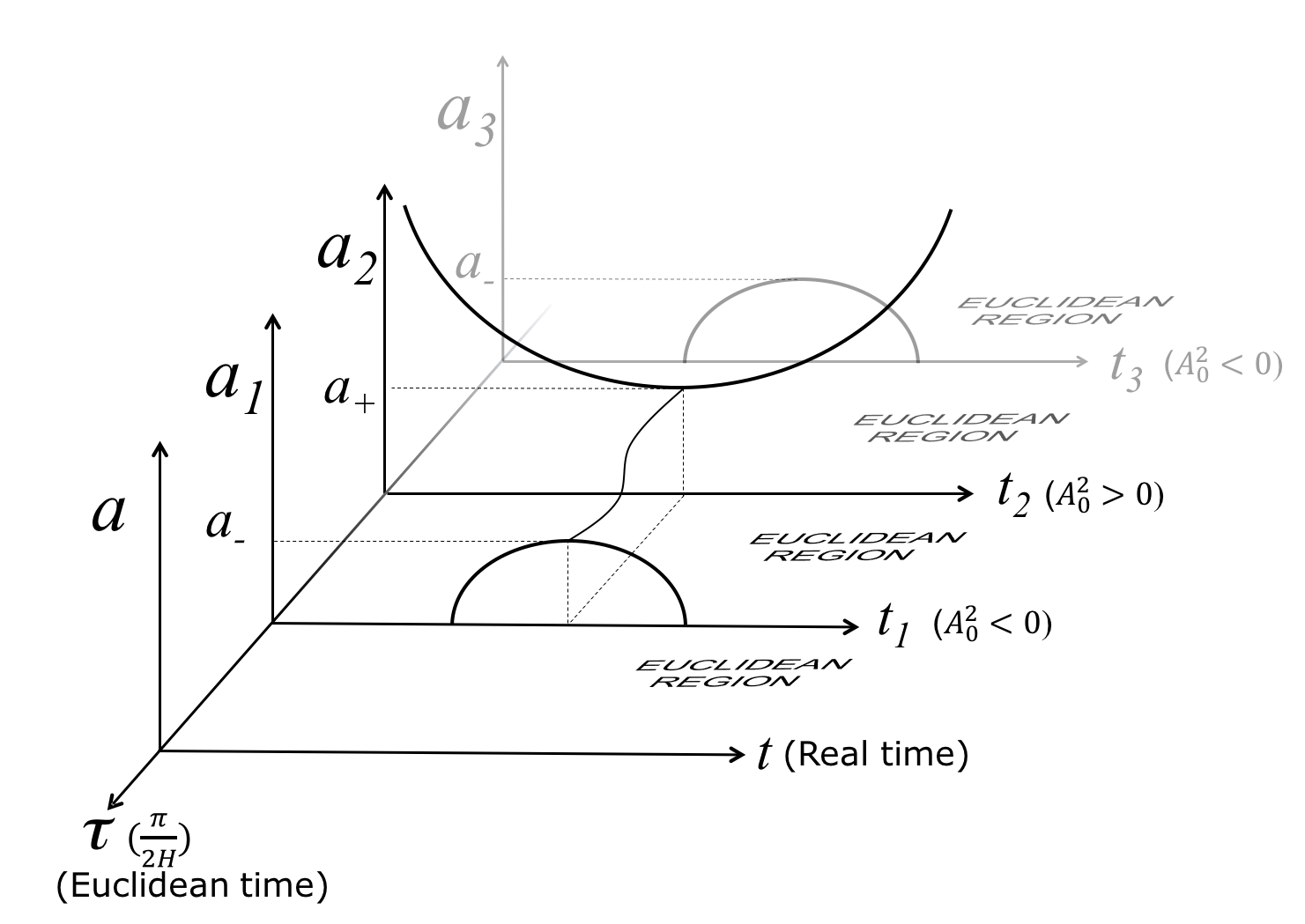}
\caption{Solution (\ref{SF12}) of the Friedmann equation for $A_0^2 = \pm 1$ and, $4 H^2 C < 1$.}
\label{figure03}
\end{figure}

\begin{figure}
\centering
\includegraphics[width=8cm]{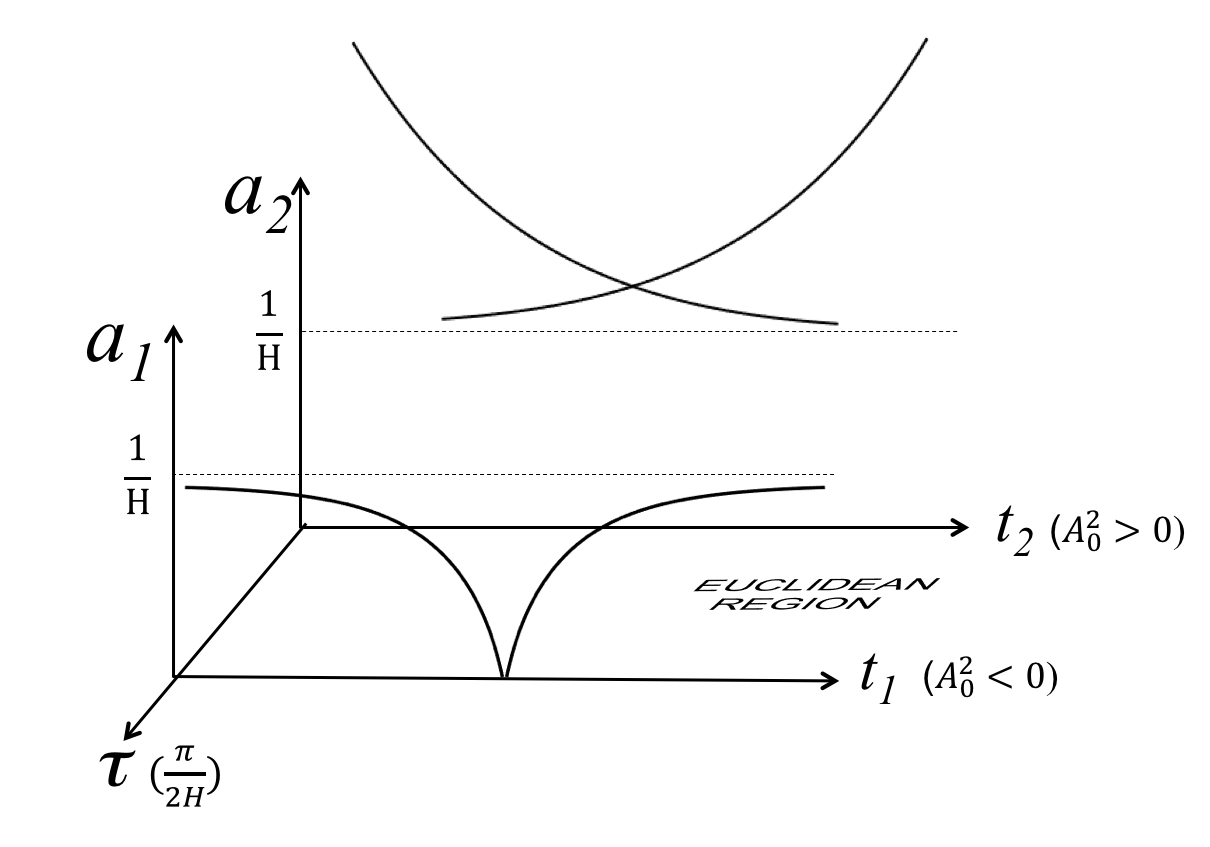}
\caption{Solution (\ref{SF13}) of the Friedmann equation for $A_0^2 = \pm 1$ and, $4 H^2 C = 1$.}
\label{figure04}
\end{figure}

The solutions of the Friedmann equation (\ref{FE02}) can easily be obtained by making the change, $u = a^2$, for which (\ref{FE02}) transforms into
\be\label{DU}
\frac{d u}{\sqrt{(u-u_+)(u-u_-)}} = 2 H t ,
\ee
where, $u_\pm \equiv a_\pm^2$. In terms of the original variable $a$, the integration of (\ref{DU}) yields
\be\label{SF01}
a^2 = \left( \frac{A_0 e^{Ht} + \frac{1}{A_0 H^2} e^{-Ht}}{2} \right)^2 - \frac{C}{A_0^2 H^2} e^{-2Ht} ,
\ee
where $A_0\equiv e^{-H t_0}$, is a constant of integration. Let us notice that $A_0$ only appears effectively in (\ref{SF01}) as $A_0^2$, so $A_0$ can generally be complex as far as $A_0^2$ is real, which is a necessary condition for the scale factor (\ref{SF01}) to be real as well. Then, let us write, $t_0 \equiv t_{0} + i \, \tau_{0}$, so that
\be\label{A02}
A_0^2 = e^{-2 H t_{0}} \left( \cos2H\tau_{0} - i \sin2H\tau_{0} \right) = \pm e^{-2 H t_{0}} ,
\ee
where it has been used that $\tau_{0}$ must satisfy
\be\label{TAU01}
\tau_{0} = \frac{n \pi}{2 H} , \ n = 0,1,2,\ldots
\ee
In particular,
\beq
A_0^2 > 0 &,& \ n = 2k , \\
A_0^2 < 0 &,& \ n = 2k+1.
\eeq
Thus, the solutions of the Friedmann equation are periodically distributed along the complex time axis (see, Figs. \ref{figure02}-\ref{figure04}). In between, there are Euclidean regions, given by 
\be
\tau_0 \in \ldots \cup (0,\frac{\pi}{2 H}) \cup (\frac{\pi}{2H}, \frac{\pi}{H}) \cup \ldots ,
\ee
that work like quantum barriers that separate the different Lorentzian regions where the classical universes live. They are thus causally disconnected from a classical point of view. We shall see however that their quantum states can be quantum mechanically correlated or entangled.

Let us now consider the solution given  by (\ref{SF01}). We can distinguish three regimes. For values $4H^2C  >1$, it can generally be written as
\be\label{SF11}
a(t) = \frac{1}{\sqrt{2}H} \left( 1 \pm (4H^2 C - 1)^\frac{1}{2}  \sinh2H\Delta t \right)^\frac{1}{2} ,
\ee
where the positive and negative signs correspond to the values, $A_0^2 > 0$ and $A_0^2 < 0$, respectively, and the real exponential of $A_0^2$ in (\ref{A02}) has been absorbed into an initial real time, $\tilde{t}_0 \in \mathbb{R}$, in $\Delta t \equiv t - \tilde{t}_0$. On the other hand, for the value $4 H^2 C < 1$, the solution (\ref{SF01}) can  be written as
\be\label{SF12}
a(t) = \frac{1}{\sqrt{2}H} \left( 1 \pm (1 - 4H^2 C)^\frac{1}{2}  \cosh2H\Delta t \right)^\frac{1}{2} ,
\ee
where here too the positive and negative signs correspond to the values, $A_0^2 > 0$ and $A_0^2 < 0$, respectively. Finally, for the particular value $4 H^2 C = 1$, the solutions of the Friedmann equation given by (\ref{SF01}) can be written as
\be\label{SF13}
a(t) = \frac{1}{\sqrt{2} H} \left( 1 \pm e^{\pm 2H \Delta t} \right)^\frac{1}{2} ,
\ee
where again the positive and negative signs in front of the exponential correspond to the values, $A_0^2 > 0$ and $A_0^2 < 0$, respectively. Let us notice that all these solutions are symmetric under a time reversal transformation because the Friedmannn equation (\ref{FE01}) is invariant under the change, $t \rightarrow - t$, so we can always distinguish two branches of each solution.

There are therefore six different solutions of the Friedmann equation that represent six different types of universes. All of them are periodically distributed along the complex time axis at points $\tau_0$ given by (\ref{TAU01}). All of them are therefore separated by Euclidean regions, i.e. classically forbidden regions, so they are all classically disconnected from the causal point of view (see Fig. \ref{figure02}-\ref{figure04}).

\subsubsection{Type I universes: $4H^2C > 1, A_0^2>0$}

In this case the solutions of the Friedmann equation are given by (\ref{SF11}) with the positive sign. These are universes that start at a big bang singularity at $t=0$ in $\Delta t \equiv t - \tilde{t}_0$, with
\be
\tilde{t}_0 = \frac{1}{2 H} \arctanh \frac{1}{2 H \sqrt{C}} ,
\ee
and expands exponentially like a flat deSitter spacetime at late times (see Fig. \ref{figure02}). It is  worth noticing that this type of universe cannot be continuously transformed into an exact ($C=0$)  deSitter universe because, $4H^2C \geq 1$.

\subsubsection{Type II universes: $4H^2C > 1, A_0^2 <0$}

These are the contracting branches of the type I universes. The scale factor is given by (\ref{SF11}) with the negative sign. They contract from a large value of the scale factor to a big crunch singularity at $t=t_{bc}$ in $\Delta t \equiv t - \tilde{t}_0$, with
\be
\Delta t_{bc}  = \frac{1}{2 H} \arctanh \frac{1}{2 H \sqrt{C}} .
\ee
They are represented in Fig. \ref{figure02}. These solutions cannot be continuously transformed either into an exact deSitter spacetime.

\subsubsection{Type III universes: $4H^2C < 1, A_0^2>0$}

The scale factor of this type of universes is given by (\ref{SF12}) with the positive sign. In terms of the values of $a_+$ and $a_-$ given in (\ref{apm}), the scale factor can be written as
\be\label{aIII}
a(t) = \left( a_+^2 \cosh^2H\Delta t - a_-^2 \sinh^2H\Delta t \right)^\frac{1}{2} ,
\ee
with, $\Delta t \in (-\infty,\infty)$. The solutions are represented in Fig. \ref{figure03}. They contract from infinity into the minimum value $a_+$, reached at $\Delta t = 0$, and they then  expand  to infinity again. This type of solution can continuously be  transformed into the customary solutions of the closed deSitter spacetime in the limit $C\rightarrow 0$. In fact, they are the natural extension of the closed deSitter universe with a radiation like energy given by $C/2$.

\subsubsection{Type IV universes: $4H^2C < 1, A_0^2<0$}

The scale factor of this type of universe is given by  (\ref{SF12}) with the negative sign. It can be written as
\be\label{aIV}
a(t) = \left( a_-^2 \cosh^2H\Delta t - a_+^2 \sinh^2H\Delta t \right)^\frac{1}{2} .
\ee
This is a universe that starts in a big-bang like singularity at $t = 0$ in $\Delta t$, with
\be
\tilde{t}_0 = \frac{1}{H} \arctanh\frac{a_-}{a_+} ,
\ee
expands to a maximum value $a_-$, at $t = \tilde{t}_0$, and then re-collapses to a big-crunch like singularity, at $t = 2 \tilde{t}_0$. For a value $H \ll 1$, the evolution of this type of universes is like that of a  radiation dominated universe. It is depicted in Fig. \ref{figure03}. In the deSitter limit, i.e. in the limit $C\rightarrow 0$, this type of solution degenerates because, $a_- \rightarrow 0$, so $\tilde{t}_0 \rightarrow 0$. It is also worth noticing that in the limit $C \rightarrow \frac{1}{4 H^2}$,  $a_- \rightarrow a_+$.

\subsubsection{Type V universes: $4H^2C = 1, A_0^2>0$}

The scale factor is now given by  (\ref{SF13}) with the positive sign, i.e.
\be
a(t) = \frac{1}{H} \left( 1 + e^{\pm 2H \Delta t} \right)^\frac{1}{2}  ,
\ee
where the positive and negative signs correspond here to the time symmetrical branches of the universe. For the positive sign, the universe expands asymptotically  like a flat deSitter universe. However, it is worth noticing that it does not start from a big bang like singularity because, 
\be
a(t\rightarrow -\infty) \rightarrow \frac{1}{H} .
\ee
The time reversal symmetric branch, i.e. that with the negative sign,  corresponds then to a universe that contracts from a very large value of the scale factor to the value $a_0 = \frac{1}{H}$ in an infinite time. Thus, it does not end in a big-crunch singularity. They are both depicted in Fig. \ref{figure04}.

\subsubsection{Type VI universes: $4H^2C = 1, A_0^2<0$}

Finally,  the scale factor of these universes is given by Eq. (\ref{SF13}) with the negative sign, i.e.
\be
a(t) = \frac{1}{H} \left( 1 - e^{\pm 2H \Delta t} \right)^\frac{1}{2}  ,
\ee
where the positive and negative signs correspond here as well to the time symmetrical branches of the universe. For the negative sign, the universe starts from a big-bang like singularity at $t = \tilde{t}_0$ and then expands to an asymptotical maximum value $a_0 = \frac{1}{H}$. It does not re-collapse. For the positive branch, the universe is created with the value $a_0$ and then re-collapses to a big-crunch singularity at $\Delta t = 0$. They are depicted in Fig. \ref{figure04}.

\begin{figure}
\centering
\includegraphics[width=8cm]{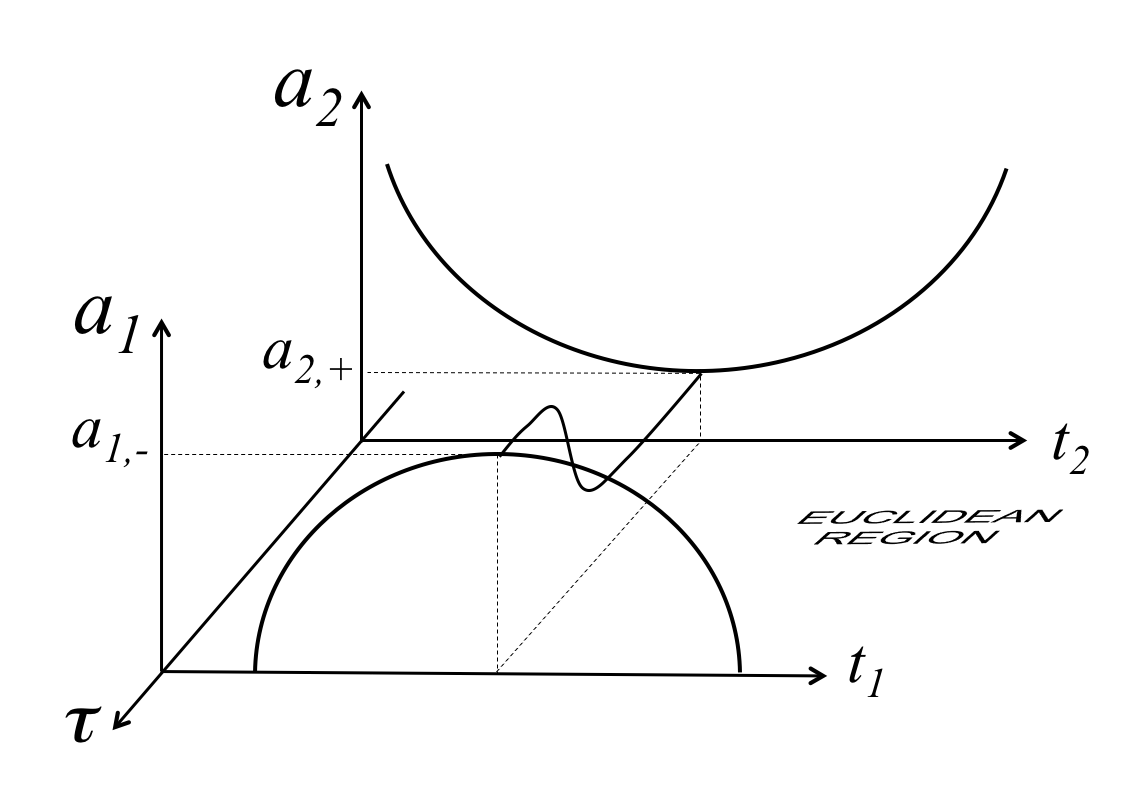}
\caption{A cyclic universe can be connected with a deSitter like universe by means of the  Euclidean instanton (\ref{INST02}).}
\label{figure05}
\end{figure}

\subsection{Euclidean solutions}

For the value $4 H^2 C \geq 1$, the r.h.s of Eq. (\ref{FE01}) is always non negative and there is thus no classically forbidden region. Lorentzian solutions given by (\ref{SF11}) or (\ref{SF13}) exist along the positive real axis, $a \in \R^+$. For the value $4 H^2 C < 1$, however, we have already pointed out that there are two Lorentzian regions separated by a Euclidean region located in the interval, $a_- < a < a_+$ (see, Fig. \ref{figure01}). We can obtain the Euclidean solutions of the Friedmann equation by Wick rotating the solution (\ref{SF01}) to Euclidean time, i.e. $t \rightarrow - i \tau$, with $A_0 \rightarrow A_E \equiv e^{\tau_{0}} e^{-i t_0}$, is a new constant of integration. In order to obtain real solutions of the Euclidean scale factor, $a_E$, $t_0$ has to satisfy now
\be
e^{2 t_0} = \pm \frac{H^2}{(1- 4 H^2 C)^\frac{1}{2}} .
\ee
On the other hand, the value of $\tau_{0}$ can be incorporated in the exponentials of (\ref{SF01}). The Euclidean solutions can then be written as
\be\label{aE}
a_E(\tau) = \left( a_+^2 \sin^2H\Delta\tau + a_-^2 \cos^2H\Delta\tau \right)^\frac{1}{2} ,
\ee
with $a_\pm$ given by (\ref{apm}), $a \in (a_-, a_+)$, and 
\be
\Delta\tau = \tau - \tau_{0} \in (0, \frac{\pi}{2 H}) .
\ee

Let us now consider the Euclidean instanton with geometry given by 
\be\label{Einst}
ds_E^2 = d\tau^2 + a_E(\tau) d\Omega_3^2 ,
\ee
with, $a_E(\tau)$ given by (\ref{aE}). The Euclidean instantons are the classical solutions of the Euclidean action. They are therefore the saddle points of the Euclidean action and provide the first order contribution to the probability of crossing the quantum barrier where they leave (see, Sect. III.B). In the case of the Euclidean instanton (\ref{Einst}), it connects the maximally expanded three sphere of a type IV universe with the minimum sphere of a type III universe, both with the same values of the cosmological constant, $\Lambda = 3 H^2$, and the same energy of the scalar field, $E = C/2$, because $H$ and $C$ are the same in $a_\pm = a_\pm(H,C)$ in (\ref{aE}).

\begin{figure}
\centering
\includegraphics[width=8cm]{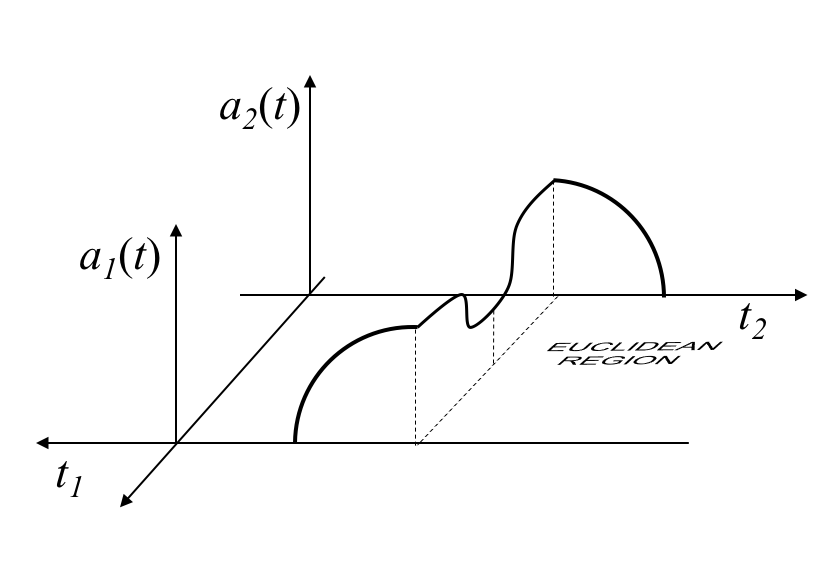}
\caption{The creation of  a pair of cyclic universes from a double instanton.}
\label{figure06}
\end{figure}

One can easily generalize the Euclidean instanton (\ref{Einst}) to another one with the same geometry and Euclidean scale factor given by
\be\label{INST02}
a_E(\tau) = \left[ a_{2,+}^2 \sin^2\bar{H}\Delta\tau + a_{1,-}^2 \cos^2\bar{H}\Delta\tau\right]^\frac{1}{2} ,
\ee
where $\Delta\tau \in (0,\frac{\pi}{2 \bar{H}})$, and
\beq\label{a1minus}
a_{1,-}^2 &=& \frac{1}{2H_1^2} \left( 1 -  \left(1 - 4 C_1 H_1^2 \right)^\frac{1}{2} \right)  ,\\ \label{a2plus}
a_{2,+}^2 &=& \frac{1}{2H_2^2} \left( 1 +  \left(1 - 4 C_2 H_2^2 \right)^\frac{1}{2} \right) .
\eeq
It is the solution of a Euclidean Friedmann equation that it is the Euclidean version of Eq. (\ref{FE01}) with effective values $\bar{H}$ and $\bar{C}$ given by
\be\label{Hbar}
\bar{H} = \frac{1}{\sqrt{a_{1,-}^2 + a_{2,+}^2}} \rightarrow H ,
\ee
and
\be\label{Cbar}
\bar{C} = \frac{a_{1,-}^2 a_{2,+}^2}{a_{1,-}^2 + a_{2,+}^2} \rightarrow C ,
\ee
where the limits are reached for the values, $H_1 = H_2 \equiv H$ and $C_1 = C_2 \equiv C$. Equivalently, this instanton is the Euclidean version of a deSitter spacetime with $\bar{\Lambda} = \frac{3 \bar{H}^2}{\sigma^2}$ and a conformally coupled massless scalar field with energy  given by $\bar{C}/2$. The instanton (\ref{INST02}) can connect the maximum expansion point of a type IV universe, given by $a_{1,-} \equiv a_-(H_1,C_1)$, with the minimum expansion point of a type III universe, given by $a_{2,+} \equiv a_+(H_2, C_2)$, regardless of the different values of the cosmological constants $H_1$ and $H_2$, and the energy density of the scalar fields, $C_1$ and $C_2$, of the two connected universes. It is worth noticing that $a_{2,+}$ can even be smaller than $a_{1,-}$. Thus, it can connect a baby universe with a parent universe like it is represented in Fig. \ref{figure03} but it can also connect a large parent cyclic universe with another parent exponentially expanding universe, like it is depicted in Fig. \ref{figure05}, or two cyclic universes (Fig. \ref{figure06}). The variety of processes resulting from the quantum transitions in this type of multiverse \cite{Marosek2015, RP2017a} generated by this type of instanton turns out to be extremely rich.

Even more, double Euclidean instantons can also connect two large cyclic universes or two large exponentially expanding universe, both with the same or with different values of their cosmological constants and the energy of their scalar fields. It is worth noticing that there is no conflict here with the energy conservation because all the solutions satisfy the Hamiltonian constraint (\ref{HAMcons}) so in that sense the total energy of all the universes is zero and the quantum transitions do not violate that condition so the total energy remains  zero in the transition. That does not mean however that all the transitions are equally probable. This will be computed in the next section.

\section{Quantum states}

\subsection{Wheeler-DeWitt equation}

Let us canonically quantize the Hamiltonian constraint (\ref{HAM02}) by promoting the dynamical variables into operators. In the configuration space  $(a,\chi)$, the Hamiltonian constraint (\ref{HAM02}) turns out to be the Wheeler-DeWitt equation, which reads \cite{Hartle1983}
\beq\nn\label{WDW01}
\frac{1}{2} \left( \frac{1}{a^p} \frac{\partial}{\partial a} \left( a^p \frac{\partial }{\partial a}\right)- a^2 + H^2 a^4 + \frac{\partial^2}{\partial \chi^2} + \chi^2 \right) & & \\ \times \Psi(a,\chi) = 0 ,
\eeq
where $p$ is a constant determining the operator ordering \cite{Hartle1983, Gott1998}, and $\Psi(a,\chi)$ is the wave function of the universe, i.e. the wave function for the composite state of spacetime and matter fields. The Wheeler-DeWitt equation (\ref{WDW01}) separates \cite{Gott1998}
\begin{eqnarray}\label{eq1}
\frac{1}{2} \left( -\frac{d^2}{d\chi^2} + \chi^2 \right) \phi(\chi) = E \phi(\chi) , \\ \label{eq2}
\frac{1}{2} \left[ -\frac{1}{a^p} \frac{d}{da}\left( a^p  \frac{d}{da} \right) +  \left( a^2 - \frac{\Lambda}{3} a^4 \right) \right] \psi(a) = E \psi(a) .
\end{eqnarray}
The first of these equations is the equation of a quantum harmonic oscillator with unit mass and frequency, which can be solved in terms of Hermite polynomials, $\mathcal{H}_n(x)$. The customary normalized solutions are
\be
\phi(\chi) = \frac{1}{\sqrt{2^n n!}} \left( \frac{1}{\pi \hbar}\right)^\frac{1}{4} e^{-\frac{\chi^2}{2 \hbar}} \mathcal{H}_n(\chi/\hbar) ,
\ee
with
\be
E \equiv E_n = \hbar (n + \frac{1}{2}) .
\ee
The second of Eqs. (\ref{eq1}-\ref{eq2}) can formally be written as the \emph{classical} equation of a harmonic oscillator
\begin{equation}\label{WDW2}
\ddot{\psi}_n(a) + \frac{\dot{\mathcal{M}}}{\mathcal{M}} \dot{\psi}_n(a) + \omega^2_n \psi_n(a) = 0 ,
\end{equation}
where the dot means derivative with respect to the scale factor, which acts then as the time like variable in (\ref{WDW2}). The \emph{time} dependent mass and frequency in (\ref{WDW2}) are given by, $\mathcal{M}\equiv\mathcal{M}(a) = a^p$, and 
\begin{equation}\label{frequency}
\omega_n\equiv\omega_n(a) = \sqrt{ H^2 a^4 - a^2 + 2 E_n} .
\end{equation}
Let us notice that for a universe for which the Wheeler-DeWitt equation is given by Eq. (\ref{WDW2}) the Friedmann equation is just given by (see Eqs. (\ref{M01}) and (\ref{M02}))
\be
\pi_a^2  = \omega_n^2 \ \text{ or}, \  p_a^2 = \omega_n^2 ,
\ee
which yields the Friedmann equation in conformal and cosmic times, Eqs. (\ref{FE00}) and (\ref{FE01}), respectively. Therefore, the frequency of the time dependent harmonic oscillator that describes the quantum state of the universe contains all the information about the evolution of the given universe. It will be especially important later on.

The general quantum state that is solution of the Wheeler-DeWitt equation (\ref{WDW01}) is therefore \cite{Hartle1983}
\be\label{WF01}
\Psi(a,\chi) = \sum_n c_n \phi_n(\chi) \psi_n(a) .
\ee
It is a superposition of composite states\footnote{In the terminology of Everett it is a superposition of relative states (see Eq. (1) of Ref. \cite{Everett1957})} where the quantum state of the matter field is completely correlated with the state of the corresponding spacetime where it propagates. On the other hand, the state $\psi_n(a)$ in (\ref{WF01}) represents the quantum state of a region of the spacetime that expands (or contracts) according to the Friedmann equation
\be\label{FE04}
\frac{d a}{dt} = \frac{\omega_n}{a} .
\ee
A given vacuum of the landscape is then populated with small regions of the spacetime that contains a the scalar field with a definite energy label, $n$, and these distinct regions of the spacetime evolve differently according to the corresponding Friedmann equation (\ref{FE04}). The quantum states of all these regions are  then entangled in a composite state (\ref{WF01}) that represents the quantum state of a simplified version of the spacetime foam \cite{Wheeler1957, Hawking1978, Garay1998}.

We have then two possibilities. In those regions where the scalar field is created in a state with a small number $n$, given by $n < n_0$, with
\be\label{n0}
n_0 \equiv \frac{1}{8 H^2} - \frac{1}{2} ,
\ee
the condition $4H^2C <1$ is satisfied and the evolution of the spacetime is given by Eq. (\ref{aIV}). A small bubble of the spacetime expands then to a maximum value $a_-$ and then re-collapses again to delve into the gravitational foam. However, at the turning point $a_-$ the state of the spacetime may undergo a quantum transition through the Euclidean region located between $a_-$ and $a_+$. Then, the bubble would  suffer a sudden transition and it would start expanding from the value $a_+$ of the scale factor to an asymptotically closed deSitter spacetime. A macroscopic universe has then been created.

On the other hand, for regions of the spacetime for which the scalar field is created with a large number $n$, given by $n > n_0$, then, $4H^2 C > 1$, and the scale factor of the spacetime in that region would follow Eq. (\ref{SF11}). Then, the bubble would start inflating from the very beginning and expanding asymptotically like a deSitter universe. Finally, in the particular case in which the number of particles would be exactly $n_0$, then, the spacetime would asymptotically expand like a  deSitter spacetime with a scale factor given by (\ref{SF13}).

One might expect that the configurations with a large number of particles of the scalar field would be exponentially suppressed and thus the cases of ever inflating solutions without tunneling transition should be exponentially forbidden. That need not necessarily be the case. First, because all possible configurations are solutions of the Wheeler-DeWitt equation and thus they all satisfy the Hamiltonian constraint, i.e. the total energy is always zero and the total energy is thus conserved in the creation of any type of universe. Secondly, because the wave function of all those configurations is regular at $a \rightarrow 0$. In that sense, it is avoided the classically singular character of the origin and the universes can be created with a very small value of the scale factor \cite{Gott1998}. Let us notice that for small values of the scale factor the wave function of the spacetime approximates that of a harmonic oscillator too, and then
\be\label{WF02}
\Psi \approx \sum_n c_n e^{-\frac{1}{2\hbar} \left( a^2 + \chi^2 \right)} \mathcal{H}_n(a/\hbar) \mathcal{H}_n(\chi/\hbar) ,
\ee
which is regular at $a = 0$. But even if the suppression of large values of $n$ would be the result of imposing a particular boundary condition on the composite state (\ref{WF01}), for instance the result of imposing the condition, $c_n = e^{-\frac{1}{T}(n+\frac{1}{2})}$ in (\ref{WF01}) or (\ref{WF02}), which would correspond to a thermal distribution of universal states at temperature $T$, even though, there would be bubbles that would inflate from the very beginning without the need of a tunneling transition. In fact, even if the scalar field is created in the lowest, $n=0$ state, there would be regions of the landscape for which $H^2 > \frac{1}{4}$. In those regions, $4H^2 C > 1$ and thus the bubbles would start exponentially inflating from the very beginning. Therefore, the solutions given by (\ref{SF11}) cannot be disregarded from the general picture.

It is also worth noticing that the superposition principle of the quantum mechanics of the matter fields alone is not satisfied here, i.e. a superposition state given by
\be\label{SPS01}
\phi(\chi) = \sum_n c_n \phi_n(\chi) ,
\ee
is not a solution of the Wheeler-DeWitt equation (\ref{WDW01}). Inside each universe, whose quantum state is represented by $\psi_n$,  the label $n$ is totally definite. There is no possibility of having superposition states like the one represented by (\ref{SPS01}). The superposition state (\ref{SPS01}) is only solution of the Hamiltonian constraint in a large macroscopic universe like ours. The superposition principle of quantum mechanics turns out to be then an emergent feature of the semiclassical state of the universe. Let us note that for large values of the scale factor we can make use of the WKB solutions of (\ref{eq2}), given by
\be
\psi_n(a) = \frac{N_n}{ \sqrt{\omega_n}} e^{\pm\frac{i}{\hbar} \int \omega_n(a) da } , 
\ee
where $N_n$ is some normalization constant, and
\be
\omega_n \approx \omega_{DS} + \frac{1}{\omega_{DS} } (n+\frac{1}{2}) ,
\ee
where $\omega_{DS}$ is the square root of the potential of a closed deSitter universe, 
\be\label{omegaDS}
\omega_{DS} = \sqrt{H^2 a^4 - a^2} .
\ee
Then, the solution of the Wheeler-DeWitt equation  can be written as
\be
\Psi(a,\chi) = \frac{N_n}{ \sqrt{\omega_n}} e^{\pm\frac{i}{\hbar} \int \omega_{DS}(a) da } \Phi(a, \chi) ,
\ee
with
\be
\Phi(\eta, \chi) \equiv \Phi(a = a(\eta), \chi) = \sum_n c_n e^{-\frac{i}{\hbar} (n+\frac{1}{2}) \eta} \phi_n(\chi) ,
\ee
where $\eta$ is the conformal time of the background deSitter spacetime, i.e
\be
\eta = \mp \int \frac{d a}{\omega_{DS}} = \mp \int \frac{dt}{a},
\ee
for the expanding ($+$) and the contracting ($-$) branches of the deSitter spacetime. The background deSitter spacetime evolves in cosmic time as,
\be\label{aDS}
a_{DS}(t) = \frac{1}{H} \cosh H t ,
\ee
which is the solution of the corresponding Friedmann equation, 
\be\label{aDS}
\frac{d a}{d t} = \frac{\omega_{DS}}{a} .
\ee
The difference between $a_{DS}(t)$ in Eq. (\ref{aDS}) and $a(t)$ in Eqs. (\ref{SF11}-\ref{SF13}) is that  the latter contain the back reaction effects of the scalar field. For a large value of the scale factor this effect is highly subdominant and the superposition principle of quantum mechanics is then satisfied, at least to order $\hbar^1$.

Of course, the superposition principle of quantum mechanics for the composite state of the matter field and the states of the spacetime is still valid,  and it is in fact at the root of the entangled state (\ref{WF01}) because it is a direct consequence of the linearity of the Wheeler-DeWitt equation (\ref{WDW01}). It is only the superposition principle of the quantum mechanics of matter fields alone what is an emergent figure of the quantum state of the universe and it  becomes valid only in a large parent universe like ours, where a time variable can be defined, but not in the composite state (\ref{WF01}) where all the regions of the spacetime are correlated and created with a definite number $n$.

\begin{figure}
\centering
\includegraphics[width=8cm]{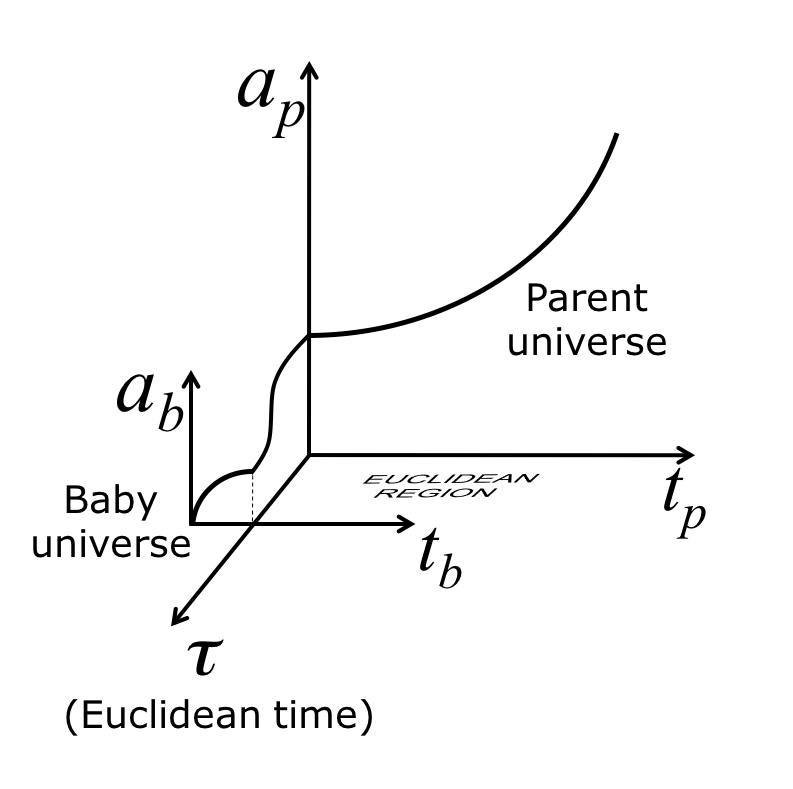}
\caption{The creation of a large parent universe from a baby universe.}
\label{figure07}
\end{figure}

\subsection{Quantum transitions}

\subsubsection{Single universe}

For the value $4H^2 C \geq 1$ there is no classically  forbidden region for any value of the scale factor. However, for the value $4H^2C < 1$, we have already pointed out that there is a Euclidean region between two Lorentzian regions that acts as a quantum barrier (see Fig. \ref{figure01}). The solutions of region $I$ in Fig. \ref{figure01} are baby universes that are created from the spacetime foam, they expand like a radiation dominated universe and at the turning point, $a = a_-$, the state of the universe can tunnel out through the Euclidean barrier and appear in region $III$ as a newborn universe with the value $a_+$ of the scale factor. The probability of the tunneling transition is given by 
\begin{equation}\label{P01}
P \propto e^{-I} ,
\end{equation}
where (see App. \ref{A2})
\begin{eqnarray} \nn
I &=& H \int_{a_-}^{a_+} da \, \sqrt{(a^2 - a_-^2)(a_+^2 - a^2)} \\ \label{I01}
&=&  -\frac{(1-4 C H^2)^\frac{1}{4}}{3 H^2 a_+^2} \left\{  C  F(q, k) - a_+^2 E(q,k) \right\},
\end{eqnarray}
where $F(q,x)$ and $E(q,x)$ are the  elliptical integrals of first and second kind, respectively, with $q \equiv \arcsin\frac{1}{k}$, and
\be
k^2 = \frac{a_+^2}{a_+^2 - a_-^2} .
\ee
The kind of transitions that can be posed in the landscape with the Euclidean instanton (\ref{INST02}) is extremely rich. First, one can pose tunneling transitions between two universes with the same value of their cosmological constant and energy of the matter fields, but one can also pose tunneling transitions between universes with different values of their cosmological constants and the energy of their scalar fields. In that case, the probability for the tunneling transition is given by, 
\be\label{P02}
P \propto e^{- I_2} ,
\ee
where $I_2$ can be written as,
\be\label{I02}
I_2 = \alpha F(q, k) - \beta E(q,k) ,
\ee
where $\alpha$ and $\beta$ are two coefficients given in (\ref{I2B}).

\subsubsection{Pair of entangled universes}

\begin{figure}
\centering
\includegraphics[width=6cm]{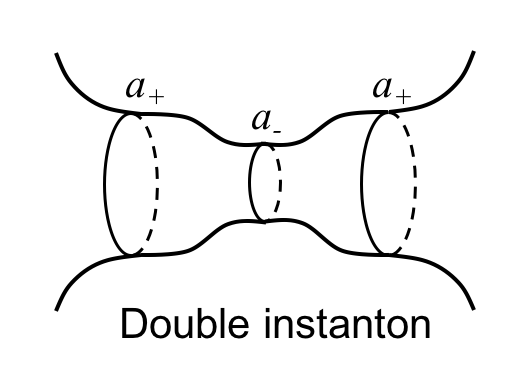}
\caption{A double Euclidean instanton can be formed by matching two single Euclidean instantons.}
\label{figure08}
\end{figure}

\begin{figure}
\centering
\includegraphics[width=7cm]{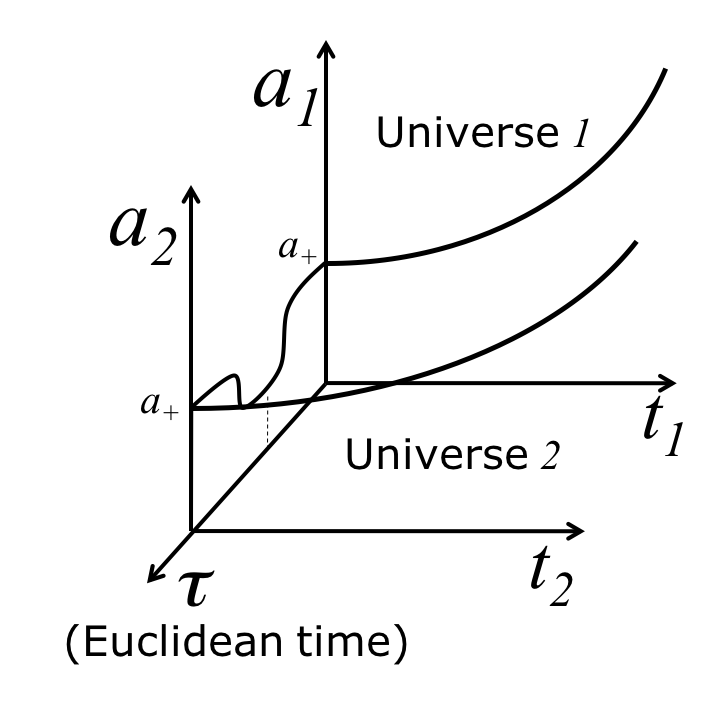}
\caption{The creation of a pair of entangled universes from \emph{nothing}, i.e. from a double Euclidean instanton.}
\label{figure09}
\end{figure}

The probabilities (\ref{P01}) and (\ref{P02}) are the probabilities for a pre-existing universe in region $I$ to tunnel out to region $III$. It is the customary picture for the creation of a universe from something \cite{Gott1998, Linde1991, RP2014}, i.e. from a pre-existing baby universe (see, Fig. \ref{figure07}). However, there is still room for the universes to be created from \emph{nothing}, i.e. from the Euclidean region without the need of a pre-existing universe. However, in order for the universes to be created from nothing they have to be created in entangled pairs \cite{RP2014, Chen2016} from a double Euclidean instanton like the one depicted in Fig. \ref{figure08} (see also Fig. \ref{figure09}). The probability for the pair of entangled universes to be created from nothing would  then be given by
\be
P = e^{-2 I} ,
\ee
where $I$ is the Euclidean action of a single instanton, given by (\ref{I01}).

\subsubsection{Multipartite entangled states}

One can even pose the creation of a multipartite $N$-entangled state like the one depicted in Fig. \ref{figure11}. In that case, the probability of having $N$-entangled universes would be given by
\be
P = e^{-N I} ,
\ee
where $I$ is given by (\ref{I01}). It means that the creation of $N$-entangled universes from a $N$-Euclidean instanton is exponentially suppressed for a large value of $N$. However, these highly non-classical states can be present as well in the multiverse.

In general, one can  pose the creation of two or more universes from a multiple Euclidean instanton formed by gluing single instantons of different types (see, Fig. \ref{figure12}). The only needed condition is that the instantons have to be matched at two hypersurfaces with the same value of the scale factor and equal tangent vector in the matching hypersurface. The latter condition is satisfied by the instanton (\ref{INST02}) at the hypersurfaces $a_+$ and $a_-$ for any value of $H$ and $C$ so the multiple instanton that can be formed is quite general (see, Fig. \ref{figure12}). The probability would eventually be given by the product of the probabilities of the single instantons, i.e.
\be
P = e^{-\sum_i N_i I_i} ,
\ee
where $N_i$ is the number of instantons of type $i$ used to form the total instanton, and $I_i$ is the Euclidean action that corresponds to each type of instanton.

\begin{figure}
\centering
\includegraphics[width=8cm]{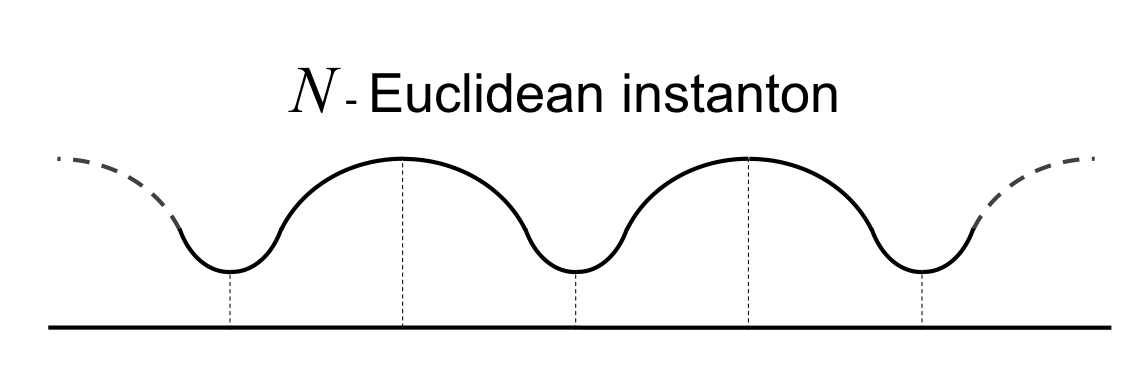}
\caption{A $N$-Euclidean instanton.}
\label{figure10}
\end{figure}

\begin{figure}
\centering
\includegraphics[width=8cm]{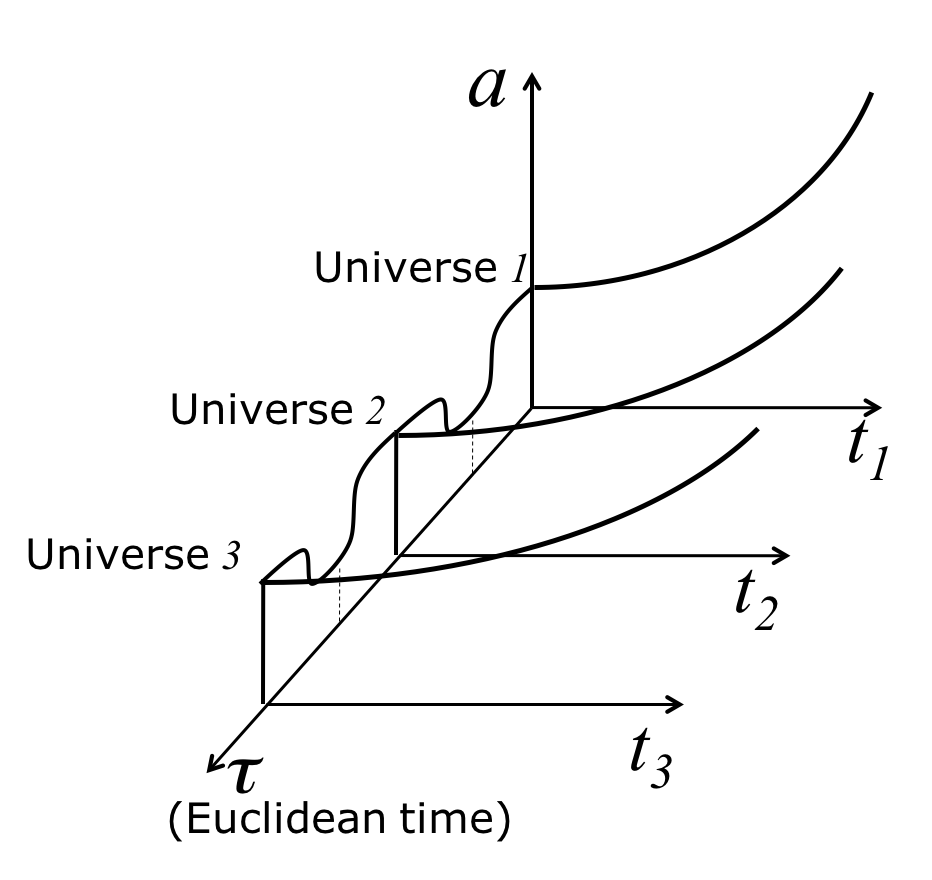}
\caption{The creation of a multipartite entangled state of $N$ universes from a $N$-Euclidean instanton.}
\label{figure11}
\end{figure}

Thus, a $N$-entangled Euclidean instanton would give rise to $N$ Lorentzian universes (see Fig. \ref{figure07}). Following the same reasoning to that made in the preceding section, the single instantons can only be matched for an equal value of the mode of the matter field and, therefore, the composite $N$-partite state of the Lorentzian regime must necessarily be an entangled state.

We can still consider the creation of a multipartite\footnote{See, for instance, Refs. \cite{Ma1990, Lo1993, Albeverio2005, Facchi2008} and references therein for the definition of multipartite entangled states in the context of quantum optics.} entangled state from both something or nothing. In the former case, a baby universe would give rise to an entangled state of $N$ universes. In the latter case, we can identify the matching hypersurfaces of the $N$ Euclidean instanton with the matching hypersurface of the first instanton, making therefore a cyclic chain of $N$-Euclidean instantons that would induce the creation from nothing of the $N$-partite entangled state in the multiverse. Furthermore, different configurations can be envisaged for the creation of $N$-entangled universe from a $N$-composite Euclidean instanton by combining both mechanism, all of them given rise to composite states of $N$ Lorentzian universes with some degree of entanglement \cite{Amico2008}.

\begin{figure}
\centering
\includegraphics[width=8cm]{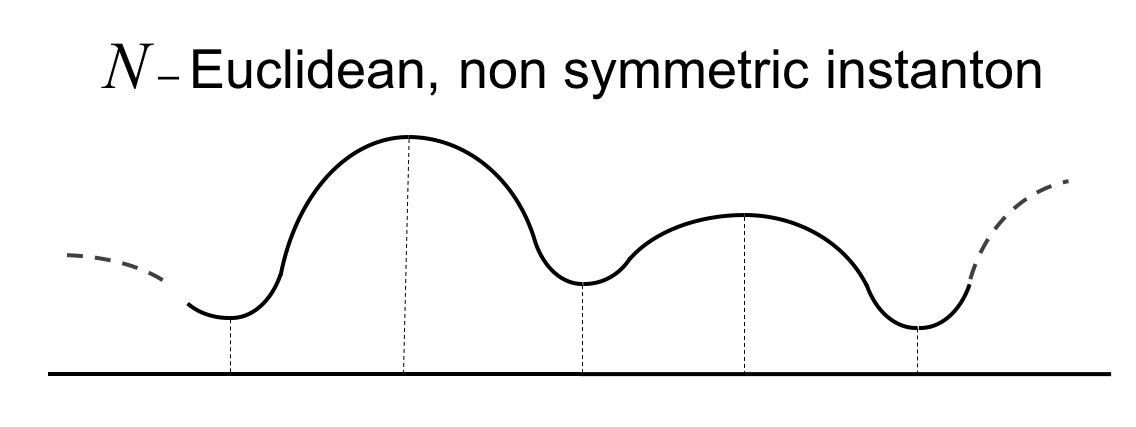}
\caption{A multiple Euclidean instanton can also be formed by gluing instantons of different types.}
\label{figure12}
\end{figure}

\section{Inter-universal entanglement}

\subsection{Third quantization formalism}

The creation and annihilation of universes is better understood in a third quantization formalism \cite{Strominger1990, RP2010}. It basically consists of considering the Wheeler-DeWitt equation (\ref{WDW01}) as the wave equation of a scalar field,  $\Psi(a,\chi)$, that propagates in the minisuperspace spanned by the configuration variables $(a,\chi)$, where the scale factor formally plays the role of the time-like variable of the minisuperspace and the scalar field $\chi$ formally plays the role of a spatial variable. Thus, the quantum state of the homogeneous and isotropic universe can be analyzed by applying a formalism that parallels that of a quantum field theory of a scalar field propagating in a curved spacetime\footnote{Let us notice that in the case considered in this paper the metric element of the minisuperspace is given by $$ds^2 = - da^2 + d\chi^2 ,$$so the minisuperspace corresponds to a Minkowski like space but in general the minisupermetric is a curved minisuperspace (see, for instance, Ref. \cite{RP2010, Garay2014}).}.

The wave function of the universe, $\Psi(a,\chi)$, is then promoted to an operator $\hat{\Psi}$ that in the Heisenberg picture is given by
\be\label{3QWF01}
\hat{\Psi}(a,\chi) = \sum_n \Psi_n(a,\chi) \hat{c}_n + \Psi^*_n(a,\chi) \hat{c}_{-n}^\dag ,
\ee
where $\hat{c}_{ n}^\dag$ and $\hat{c}_{ n}$ are constant operators that represent  the creation and annihilation of universes, respectively,  whose quantum mechanically states are described by the wave function, $\Psi_n(a,\chi) = \psi_n(a) \phi_n(\chi)$, i.e. they create or annihilate branches of the universe with a definite energy level of the scalar field $\chi$, given by $n$, and a homogeneous and isotropic geometry given by (\ref{MT01}) with a scale factor evolving according to (\ref{FE01}). The positive and negative values in $\hat{c}_n$ and $\hat{c}_{-n}^\dag$ in (\ref{3QWF01}) refer to the expanding and the contracting branches that they create or annihilate. Let us notice that the Friedmann equation is given by the Hamiltonian constraint $p_a^2 = \omega^2$ with the classical value, $p_a = - a \frac{d a}{d t}$. Then
\be
\frac{d a}{d t} = \pm \frac{\omega}{a} ,
\ee
is the Friedmann equation that corresponds, with respect to the cosmic time $t$, to a contracting ($-$) or an expanding ($+$) branch of the universe. Let us recall that for every classical solution its  time reversal transformation is a solution as well because the invariance of the Friedmann equation (\ref{FE02}) with respect to a time reversal change, $t\rightarrow - t$. Let us also notice that these two solutions are quantum mechanically  represented by the complex conjugated pair of WKB solutions
\be\label{Psin01}
\Psi_n(a,\chi) = \frac{N_n}{\sqrt{a \, \omega_{DS}}} e^{\pm\frac{i}{\hbar} S_{DS}(a)} \phi_n(a,\chi) , 
\ee
where $N_n$ is a normalization constant, $\omega_{DS}$ is given by (\ref{omegaDS}),
\be
S_n(a) = \int^{a} da' \ \omega_{DS}(a') = \frac{1}{3H^2} \left( H^2 a^2 - 1\right)^\frac{3}{2},
\ee
and, to order $\hbar^0$,
\be
p_a = \pm \frac{\partial S_n}{\partial a} = \pm \omega_n .
\ee
Each branch represents a universe and the universes are then created in entangled pairs. It parallels the creation of particles in entangled pairs in a quantum field theory. In that case, the particles are created in entangled pairs with opposite values of their momenta, $\pm k$, because the isotropy of the background spacetime and because conservation of the total momentum. Here, the universes are created as well in entangled pairs with opposite momenta, given by $p_a = \pm \omega_n$. The momentum conjugated to the scale factor is however related to the Friedmann equation and they thus correspond to reversely evolving branches of the spacetime. However, it is worth noticing that the consideration of the spacetime as expanding or contracting in this context depends on the time variable chosen by a particular observer that inhabit one of the branches, for which his or her branch is expanding and the opposite one is then contracting. For an observer living in the partner branch the situation is the other way around, his or her universe is the expanding branch and the opposite one is the contracting branch. The WKB time for these two observers turn out to be related by an antipodal like symmetry \cite{Linde1994, Linde1991}. Let us notice that from the standpoint of quantum cosmology time is an emergent feature and not the other way around. To see it, let us introduce the WKB solutions (\ref{Psin01}) into the semiclassical limit of the Wheeler-DeWitt equation. For a large parent universe like ours it is satisfied
\be
\pm 2 i \hbar \omega_{DS} \frac{\dot{\phi}}{\phi} - \hbar^2 \frac{\phi''}{\phi} + \chi^2 = 0,
\ee
which is equivalent to the time dependent Schr\"{o}dinger equation
\be\label{SCH01}
-i \hbar \frac{\partial }{\partial t} \phi(t,\chi) = \frac{1}{2} \left( -\hbar^2 \frac{\partial^2}{\partial \chi^2} + \chi^2 \right) \phi(t,\chi) ,
\ee
where, $\phi(t,\chi) \equiv \phi(a(t),\chi)$, and the time variable $t$ has been defined according to 
\be
\frac{\partial }{\partial t} = \pm \ \frac{\omega_{DS}}{a} \frac{\partial}{\partial a } .
\ee
Therefore, from the quantum mechanical standpoint time arises as an emergent feature of the semiclassical regime of the wave function of the universe. An observer inhabiting each branch of the entangled pair defines his or her time variable from the experiments of particle physics that are governed by the Schr\"{o}dinger equation (\ref{SCH01}). Thus the two WKB solutions describe both an expanding universe from the point of view of the time variable experienced by an internal observer (see Fig. \ref{figure09}).

The third quantization formalism parallels that of a quantum field theory propagating in a curved spacetime. It is only valid to describe universes with high degree of symmetry, but this is enough to describe most of the evolution of a universe like ours. A similar effect happens in a generally curved spacetime where is not always possible to define a consistent time variable and thus, a well-defined quantum field theory cannot be developed. In the case of the multiverse the scale factor formally plays the role of the time like variable of the minisuperspace of geometries and matter fields and the latter formally play the role of the spatial variables. Then, for homogeneous and isotropic spacetimes the third quantization formalism is well defined.

We have now to choose the appropriate vacuum state of the wave function $\Psi(a,\chi)$ in the minisuperspace by imposing a particular boundary condition. For this, we impose that the vacuum state must be a stable vacuum state and it thus must steadily represent the ground state of the wave function of the multiverses for any value of the scale factor.

Following the prescription of a quantum field theory the vacuum state would be given by the composite state
\be
|0 \rangle_b = \prod_n |0_n \rangle_b ,
\ee
where $|0_n\rangle_b$ is the ground state of an invariant annihilation operator, $\hat{b}_n$, for each mode of the composite state (\ref{3QWF01}), i.e.
\be
\hat{b}_n | 0_n \rangle = 0.
\ee
The invariant operators $\hat{b}_n$ and $\hat{b}_n^{\dag}$ are such that
\be
\frac{d \hat{b}_n^{(\dag)}}{d a } = \frac{-i}{\hbar} [\hat{H}_n ,  \hat{b}_n^{(\dag)}] + \frac{\partial \hat{b}_n^{(\dag)}}{\partial a} = 0 ,
\ee
where $\hat{H}_n$ is the third quantized Hamiltonian for which the Heisenberg equations give rise to the Wheeler-DeWitt equation (\ref{WDW2}), i.e. 
\be\label{H301}
\hat{H}_n = \frac{1}{2 \mathcal{M}} P_\psi^2 + \frac{\mathcal{M} \omega^2}{2} \psi^2 ,
\ee
with $\mathcal{M}$ and $\omega_n^2$ being given after (\ref{WDW2}). Then, the eigenstates $|N, a \rangle_b$ of the number operator of an invariant representation, $\hat{N}(a)$, have the great advantage that they are stable under the evolution of the universe because
\be
\hat{N}_n(a) |N, a \rangle_b = N |N, a \rangle_b ,
\ee
with, $N \neq N(a)$, being a constant. In particular, 
\be
\hat{N}_n(a) |0\rangle_b = 0 ,
\ee
along the entire evolution of the wave function $\Psi(a,\chi)$. It is thus a steady ground state that can represent the no universe state at any value of the scale factor.

An invariant representation of the generalized harmonic oscillator (\ref{WDW2}) can be written as\footnote{For the invariant representations of the harmonic oscillator, see for instance, Refs. \cite{Lewis1969, Leach1983, Dantas1992, Kanasugi1995, Park2004}.}
\begin{eqnarray}\label{IR01}
\hat{b}_n &=& \sqrt{\frac{1}{2}} \left( \frac{1}{R} \hat{\psi} + i(R \hat{P}_\psi- \mathcal{M} \dot{R} \hat{\psi} ) \right) , \\ \label{IR02}
\hat{b}^\dag_{-n} &=& \sqrt{\frac{1}{2}} \left( \frac{1}{R} \hat{\psi} - i(R \hat{P}_\psi -  \mathcal{M}  \dot{R} \hat{\psi} ) \right) ,
\end{eqnarray}
where $\hat{\psi}$ and its conjugated momentum $\hat{P}_\psi$ are constant operators in the Schr\"{o}dinger picture, and $R\equiv R(a)$ is an auxiliary real function that satisfies
\be\label{DDR01}
\ddot{R}_n + \frac{\dot{\mathcal{M}}}{\mathcal{M}} \dot{R}_n + \omega^2_n R_n = \frac{1}{\mathcal{M}^2 R^3} .
\ee
The invariant representation (\ref{IR01}-\ref{IR02}) must be fixed by imposing the asymptotic condition that for a large parent universe it approaches the diagonal representation given by
\beq\label{DR01}
\hat{c}_n &=& \sqrt{\frac{\mathcal{M} \omega}{2}}  \left(  \hat{\psi} + \frac{i}{\mathcal{M} \omega}  \hat{P}_\psi \right), \\  \label{DR02}
\hat{c}^\dag_{-n} &=& \sqrt{\frac{\mathcal{M} \omega}{2}} \left( \hat{\psi} - \frac{i}{\mathcal{M} \omega} \hat{P}_\psi  \right) ,
\eeq 
in terms of which the branches represent two independent universes. It means that the two branches are originally created in an entangled pair and the entanglement is decreasing as the universes expand and become large parent universes like ours. However, the effects of their entanglement could be significant in the very early stages of their evolution and they might even have a residual effect in our current universe.

The entanglement rate can be seen by writing the Hamiltonian (\ref{H301}) in terms of the invariant representation (\ref{IR01}-\ref{IR02}). It describes then the evolution of  two interacting universes whose interaction is decreasing as the universes expand. Let us notice that the invariant and the diagonal representations, given by (\ref{IR01}-\ref{IR02}) and (\ref{DR01}-\ref{DR02}) respectively, are related by the Bogolyubov transformation 
\begin{eqnarray}\label{BO01}
\hat{b}_{-n} &=& \alpha \hat{c}_{-n} - \beta \hat{c}_n^\dag , \\ \label{BO02}
\hat{b}_{-n}^\dag &=& \alpha^* \hat{c}_{-n}^\dag - \beta^* \hat{c}_n ,
\end{eqnarray}
where
\begin{eqnarray}
\alpha_n &=& \frac{1}{2} \left( \frac{1}{R\sqrt{\mathcal{M} \omega_n}} + R \sqrt{\mathcal{M} \omega_n} - i\sqrt{\frac{ \mathcal{M}}{\omega_n}} \dot{R} \right) , \\
\beta_n &=& - \frac{1}{2} \left( \frac{1}{R\sqrt{ \mathcal{M}  \omega_n}} - R \sqrt{\mathcal{M}  \omega_n} - i\sqrt{\frac{ \mathcal{M}}{\omega_n}} \dot{R} \right) .
\end{eqnarray}
Then,
\be
H = \sum_n \hbar \omega_n (\hat{c}_n^\dag \hat{c}_n + \frac{1}{2}) + \hbar \omega_n (\hat{c}_{-n}^\dag \hat{c}_{-n} + \frac{1}{2})
\ee
turns out to be 
\be\label{HI02}
H = H_1 + H_2 + H_I ,
\ee
with
\beq
H_{(1,2)} &=& \sum_n \Omega_n(a) ( \hat{b}^\dag_{(1,2), n} \hat{b}_{(1,2),n} + \frac{1}{2}) , \\
H_I &=& \sum_n \gamma_n(a) \hat{b}^\dag_{1,n} \hat{b}^\dag_{2,n} + \gamma_n^*(a) \hat{b}_{1,n} \hat{b}_{2,n} ,
\eeq
where,
\beq
\Omega_n(a) &=& \frac{1}{2} \left( \frac{1}{\mathcal{M} R^2} + \mathcal{M} R^2 \omega_n^2 + \mathcal{M} \dot{R}^2 \right)  , \\
\gamma_n(a) &=&  \Omega_n(a) -\frac{1}{\mathcal{M} R^2} + i \frac{\dot{R}}{R} .
\eeq
The Hamiltonian (\ref{HI02}) is the Hamiltonian of two interacting universes with a Hamiltonian of interaction given by $H_I$. Let us notice that a solution of (\ref{DDR01}) can generally be given by\footnote{For more general solutions, see Ref. \cite{Leach1983}.}
\be
R = \sqrt{\psi_1^2 + \psi_2^2} ,
\ee
where $\psi_1$ and $\psi_2$ are two particular solutions of the wave equation (\ref{WDW2}). In the WKB regime one can choose the following two real solutions
\beq
\psi_1 &=& \frac{1}{\sqrt{\mathcal{M} \omega_n}} \cos S , \\
\psi_2 &=& \frac{1}{\sqrt{\mathcal{M} \omega_n}} \sin S ,
\eeq
so that
\be
R =  \frac{1}{\sqrt{\mathcal{M} \omega_n}} .
\ee
This value of $R$ in (\ref{IR01}-\ref{IR02}) fulfills the boundary condition that for a large parent universe the invariant representation approaches the diagonal representation, i.e.
\be
\hat{b}_n \rightarrow \hat{c}_n \ , \ \hat{b}_{-n}^\dag \rightarrow \hat{c}_{-n}^\dag  ,
\ee 
for, $a \gg 1$. Therefore, the entanglement rate between the two universes disappears as the universes expand and become large parent universes because
\be
|\gamma_n(a) | = \frac{\xi(a)}{2 \omega_n} \sqrt{1 + \frac{1}{4} \xi^2(a)} \rightarrow \mathcal{O}(V^{-1}),
\ee
where, $V(a) = a^3$, and
\be
\xi(a) = \frac{\dot{M}}{M} + \frac{\dot{\omega}_n}{\omega_n} \sim \frac{1}{a} .
\ee
For the earliest stage of the evolution of the universes the entanglement between their quantum states can be significant and it may have an important effect of their evolution. Let us notice that the effective value of the Friedmann equation would be given by the Friedmann equation (\ref{FE04}) with the effective value of the frequency given by $\Omega_n(a)$ instead of $\omega_n(a)$. Then, at large enough values of the scale factor
\be\label{FE06}
\frac{d a}{d t} = \frac{\omega_n}{a} \left(  1 + \frac{\xi^2(a)}{8 \omega_n^2} \right) \sim H a  \left(  1 + \frac{1}{8 H^2 V^2}   \right) .
\ee
The extra term in the effective Friedmann equation (\ref{FE06}) induces a modification in the pre-exponential  stage of the evolution of the universe that might leave observable imprints in the power spectrum of the CMB \cite{Bouhmadi2011, Scardigli2011} provided that the inflationary stage does not last for too long.

\subsection{Entanglement between the scalar field of the two branches}

For a large parent universe the gravitational degrees of freedom are frozen out and the semiclassical state of the universe is described by the wave function  (\ref{Psin01}). It represents an exponentially expanding deSitter background spacetime where the field propagates and behaves quantum mechanically acoording to the Schr\"{o}dinger equation (\ref{SCH01}), which is the Schr\"{o}dinger equation of a harmonic oscillator with unit frequency and mass whose wave equation is then given by (\ref{WE01}). The second quantization of the scalar field would follow the customary procedure of promoting the scalar field to an operator that, in the Heisenberg picture, can be written as\footnote{Note however that so far we are just dealing  with the homogeneous and isotropic mode of the conformally coupled massless scalar field.}
\be\label{ENT01}
\hat{\chi}(\eta) = \chi(\eta) \hat{c}_1 + \chi^*(\eta) \hat{c}_2^\dag .
\ee
However, the creation of the universes in entangled pairs makes that their scalar fields are entangled too. Then, creation and annihilation operators, $\hat{c}_{1,2}^\dag$ and $\hat{c}_{1,2}$ respectively, in (\ref{ENT01}) turns out to be the creation and annihilation operators of particles of the scalar field in the universes $1$ and $2$. The amplitude $\chi(\eta)$ satisfies the wave equation (\ref{WE01}), whose solution is simply given in conformal time by
\be\label{n0SOL}
\chi(\eta) = e^{- i \eta} .
\ee
The composite vacuum state of the scalar field $\chi$, $|0_1 0_2 \rangle$, is annihilated by the annihilation operators $\hat{c}_1$ and $\hat{c}_2$. It is a stable vacuum state because the wave equation (\ref{WE01}) is the wave equation of a time independent harmonic oscillator. Then, if the scalar field $\chi$ is in the vacuum state at a given initial moment $\eta_0$ it will stay in that vacuum state along the entire evolution of the field. It thus represents the no particle state for all time. Let us also notice that in terms of the invariant representation of the scalar field $\hat{\chi}$, given by
\be
\hat{\chi}(\eta) = \hat{c}_1(\eta) + \hat{c}_2^\dag(\eta) \ , \ \pi_\chi = -i ( \hat{c}_1(\eta) - \hat{c}_2^\dag(\eta) ) ,
\ee
the Hamiltonian
\be
H = \frac{1}{2} \hat{\pi}_\chi^\dag \hat{\pi}_\chi + \frac{1}{2} \hat{\chi}^\dag \hat{\chi} ,
\ee
turns out to represent two non interacting scalar fields, i.e. 
\be
H = H_1 + H_2 ,
\ee
each one representing the independent evolution of the scalar field in each single universe of the entangled pair, with
\be
H_{1,2} = \hat{c}_{1,2}^\dag \hat{c}_{1,2} + \frac{1}{2} .
\ee
In that sense, the scalar fields of the two universes are not entangled. Let us notice that although the quantum states of the background spacetimes are entangled, the scalar field $\chi$ is decoupled from the spacetime degrees of freedom and thus, the modes of $\chi$ in one universe are unentangled from the modes of $\chi$ in the partner universe.

However, in terms of the original scalar field $\varphi(t)$ (see Eq. (\ref{VPHI01})), the modes of the scalar field in the entangled pair of universes are entangled too. Let us notice that in a homogeneous and isotropic universe the spatial modes are decoupled to each other and the action of the $n$ mode reads, in terms of cosmic time $t$, 
\be
S_n  = \frac{\sigma^2}{2}  \int dt \ a^3 \left( \dot{\varphi}_n^2 - \nu_n^2(t) \varphi_n^2  \right) ,
\ee
with
\be\label{FRn01}
\nu_n^2(t)  = \frac{\dot{a}^2}{a^2} + \frac{\ddot{a}}{a} + \frac{n^2 - 1}{a^2} ,
\ee
evaluated at $a(t)$. For $n=1$ it is recovered the solution (\ref{n0SOL}) of the scalar field $\chi(\eta)$ in terms of conformal time. For other modes, the wave equation for $\chi_n(\eta)$ is given by
\be
\chi_n''(\eta) + n^2 \chi_n(\eta) = 0,
\ee
whose solutions are
\be
\chi_n(\eta) = \frac{1}{\sqrt{n}} e^{- i n |\eta|} ,
\ee
which are well defined for all time. However, in terms of the cosmic time and using the Friedmann equation (\ref{FE01}), the frequency (\ref{FRn01}) yields
\be\label{FRn02}
\nu_n^2(t) = 2 H^2 + \frac{n^2 -1}{a^2} .
\ee
The momentum conjugated to the scalar field $\varphi_n$ is  given by
\be
p_\varphi = a^3 \dot{\varphi}_n ,
\ee
and the canonical transformation that relates $\chi_n$, $\pi_\chi$ with $\varphi_n$, $p_\varphi$ is
\beq
\chi_n &=& \sigma a \varphi_n , \\
\pi_\chi &=& \frac{1}{\sigma a} p_\varphi + \sigma a \dot{a} \varphi_n .
\eeq
Let us then consider the invariant representation of the modes $\chi_n$
\beq
\hat{\chi}_n &=& \frac{1}{\sqrt{n}} ( \hat{c}_{1,n} + \hat{c}_{2,n}^\dag ) , \\
\hat{\pi}_\chi &=& - i \sqrt{n} (\hat{c}_{1,n} - \hat{c}_{2,n}^\dag ) ,
\eeq
and the diagonal representation of the modes $\varphi_n$,
\beq
\hat{\varphi}_n &=& \frac{1}{\sqrt{M \nu_n}} (\hat{b}_{1,n} + \hat{b}_{2,n}^\dag) , \\
\hat{p}_\varphi &=& - i \sqrt{M \nu_n} (\hat{b}_{1,n} - \hat{b}_{2,n}^\dag) ,
\eeq
with, $M = \sigma^2 a^3$ and $\omega_n$ given by (\ref{FRn02}). Then, 
\beq
\hat{c}_{1,n} &=& \alpha_n \hat{b}_{1,n} - \beta_n^* \hat{b}_{2,n}^\dag , \\
\hat{c}_{2,n}^\dag &=& \alpha_n^* \hat{b}_{2,n}^\dag - \beta_n \hat{b}_{1,n} ,
\eeq
where
\beq
\alpha_n &=& \frac{1}{2} \left( \frac{1}{\sigma a} \sqrt{\frac{M \nu_n}{n}} + \sigma a \sqrt{\frac{n}{M \nu_n}}  + \frac{i \sigma a \dot{a}}{\sqrt{M \nu_n n}} \right) , \\
\beta_n &=&  \frac{1}{2}  \left(  \frac{1}{\sigma a} \sqrt{\frac{M \nu_n}{n}} - \sigma a \sqrt{\frac{n}{M \nu_n}}  + \frac{i \sigma a \dot{a}}{\sqrt{M \nu_n n}} \right) ,
\eeq
with, $|\alpha_n|^2 - |\beta_n|^2 = 1$, for all $n$. It turns out that the invariant vacuum state $|0_1 0 _2 \rangle_c$, given now by the tensor product
\be
|0_1 0_2\rangle_c = \prod_n |0_{1,n}\rangle_c |0_{2,n}\rangle_c
\ee
can  be related to the number states of the $b$, $b^\dag$ representation as
\be\label{VS01}
| 0_1 0_2\rangle_c = \prod_n \frac{1}{|\alpha|} \sum_{m=0}^\infty \left( \frac{|\beta_n| }{ |\alpha_n|} \right)^m |m_{1,n} m_{2,n}\rangle_b .
\ee
The reduced density matrix that describes the quantum states of the scalar field $\varphi(t)$ in one of the universes is then \cite{Mukhanov2007, RP2012}
\be\label{RHO01}
\rho_1 = \text{Tr}_2 \rho = \prod_n \frac{1}{|\alpha_n|^2} \sum_m \left( \frac{|\beta|}{|\alpha|} \right)^{2m}  |m_{1,n}\rangle_b {}_b\langle m_{1,n} | ,
\ee
which is a quasi thermal state that can be written as
\be\label{RHO02}
\rho_1 = \prod_n \rho_{1,n} = \prod_n \frac{1}{Z_n} \sum_m e^{-\frac{1}{T_n}(m+\frac{1}{2})} |m_{1,n}\rangle_b {}_b\langle m_{1,n} | ,
\ee
where, $Z_n^{-1} = 2 \sinh\frac{1}{2T_n}$, and the specific  temperature of entanglement for each mode  given by
\be\label{T01}
T_n \equiv T_n(t) = \frac{1}{ \ln \frac{|\alpha_n|^2}{|\beta_n|^2}} .
\ee
The \emph{specific} temperature (\ref{T01}) is defined as the temperature of entanglement of the state (\ref{RHO02}) per unit of frequency, 
\be
T_n \equiv \frac{\bar{T}_n}{\nu_n} ,
\ee
where $\bar{T}_n$ is the temperature of the entanglement of the quasi thermal distribution (\ref{RHO02}). However, the \emph{thermal} character of the distribution (\ref{RHO02}) is not clear because it is derived from highly non-local correlations unlike the customary definition of a thermal state in classical thermodynamics, which is obtained from the average of microscopic contact effects. In fact, the relationship between the thermodynamics of entanglement and the classical formulation of thermodynamics is a subject of intense research \cite{Plenio1998, Horodecki2008, Brandao2008, Hossein2008, Jennings2010, Cwiklinski2015}. To our knowledge there is no conclusive result yet. Besides, we are interested here in the weight of each mode in the distribution (\ref{RHO02}) and therefore, we are interested in the value of  $T_n$ rather than that of $\bar{T}_n$.

\begin{figure}
\centering
\includegraphics[width=9cm]{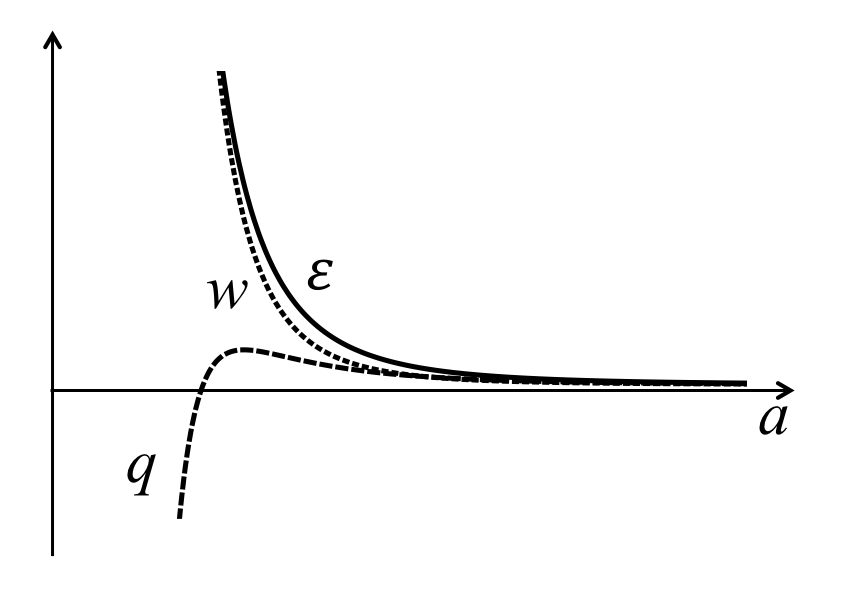}
\caption{Energy ($\varepsilon$), work ($w$), and heat ($q$) densities of entanglement, Eq. (\ref{edensity}).}
\label{figure13}
\end{figure}

Let us now consider the function $F_n(t)$ given by
\be
F_n(t) \equiv \frac{|\alpha|^2}{|\beta|^2} \equiv \frac{G_n(t) + 2}{G_n(t) - 2} ,
\ee
where,
\be\label{G01}
G_n(t) = \frac{3 H^2 a^4 + 2(n^2 - 1) a^2 + C}{a^2 n \sqrt{2 H^2 a^2 + n^2 - 1}} .
\ee
In obtaining (\ref{G01}) it has been used (\ref{FRn02}) and the Friedmann equation (\ref{FE02}). In the limit of large modes, $n \rightarrow \infty$, 
\be
G_n(t) \rightarrow 2 , \ \text{and} \ F_n(t) \rightarrow \infty \ \forall n .
\ee
Thus, $T_n \rightarrow 0$ in (\ref{RHO02}) and  
\be
\rho_{1,n} \approx   |0_{1,n}\rangle_b {}_b\langle 0_{1,n} | , \forall n\gg 1 .
\ee
Therefore, the highest modes of the scalar field stay in the vacuum state. It means that they are unaware of the effects of the entanglement between the universes. However, the lowest modes of the scalar field are heavily affected by the inter-universal entanglement. Let us notice that in the limit of large values of the scale factor, 
\be
G_n(t) \rightarrow \frac{3\sqrt{2}}{2} \frac{H a}{n} \rightarrow \infty .
\ee
Then, 
\be\label{T02}
T_n(t) \approx \frac{3\sqrt{2}}{8} \frac{H a}{n} \rightarrow \infty .
\ee
It means that the excites states of the modes are distributed with almost the same probability. The lowest modes can  then be highly excited and, thus, the effect of the entanglement between universes is expected to be large for these modes. In terms of the physical length defined as, $L_\text{ph} = \frac{a}{n}$, the temperature of the modes (\ref{T02}) turns out to be
\be
T(L_\text{ph}) \sim \frac{L_\text{ph}}{H^{-1}} .
\ee
Therefore, at short distances with respect to the Hubble length, $L_\text{ph} \ll H^{-1}$, the modes are in the vacuum states and thus the effects of the entanglement between the universes is negligible. However, at distances of order of the Hubble length, $ L_\text{ph} \gtrsim H^{-1}$, the effects becomes significant, a result that agrees with the one obtained in Ref. \cite{Mersini2008}.

Let us also notice that the state (\ref{RHO02}) can be interpreted in terms of particle creation, the number of which would be given by
\be\label{N01}
N_n = \frac{1}{e^\frac{1}{T_n} - 1} .
\ee
For large modes, $N_n \rightarrow 0$, and there is no particle creation. However, for small modes the number of particles would be
\be
N_n(t) \propto T_n(t) \sim \frac{1}{n} e^{H t} ,
\ee
a result that can be related with that obtained by Grishchuk and Sidorov\footnote{They however studied the modes of the gravitational waves instead of the modes of a scalar field.} \cite{Grishchuk1990}. One could even define the thermodynamical magnitudes of entanglement associated to the quasi thermal state (\ref{RHO02}). They are given, for each mode, by \cite{RP2012}
\beq\label{Eent}
E_n(t) &=& \frac{\omega_n}{2} \cotanh\frac{1}{2 T_n} , \\
Q_n(t) &=& \frac{\omega_n}{2} \cotanh\frac{1}{2 T_n} - \omega_n T_n \ln\sinh\frac{1}{2 T_n}, \\
W_n(t) &=& \omega_n T_n \ln\sinh\frac{1}{2 T_n} ,
\eeq
with, $E_n(t) = Q_n(t) + W_n(t)$, for all modes $n$. In Fig. \ref{figure13} it is depicted the energy densities that correspond to $E_n$, $Q_n$, and $W_n$, given respectively by
\be\label{edensity}
\varepsilon_n = \frac{E_n}{V} \ , \ q_n = \frac{Q_n}{V} \ , \ w_n = \frac{W_n}{V} ,
\ee
with, $V = a^3(t)$. They vanish for a large value of the scale factor. However, they might contribute significantly during the early phases of the evolution of the universe, with an effective value of the Friedmann equation given by
\be\label{FE08}
\frac{d a}{d t} = \frac{\omega}{a} + \frac{\xi^2(a)}{8 a \omega(a)} + a\sqrt{\varepsilon(a)} ,
\ee
with
\be
\varepsilon(a) = \sum_n^{n_\text{max}} \varepsilon_n(a),
\ee
where the sum of the modes is extended to some maximum mode that for subhorizon modes would be given by \cite{Mersini2008} $n_\text{max}  \sim a H$ (i.e., $L_\text{ph} \sim H^{-1}$). The first term in (\ref{FE08}) corresponds to the unperturbed Friedmann equation. The second term is the correction to the Friedmann equation due to the entanglement between the spacetimes of the entangled universes, and the last term is the correction originated by the entanglement between the modes of the scalar field of the two entangled universes. The effective value of the expansion rate, $H \equiv \frac{1}{a} \frac{da}{dt}$, is depicted in Fig. \ref{}.

\begin{figure}
\centering
\includegraphics[width=9cm]{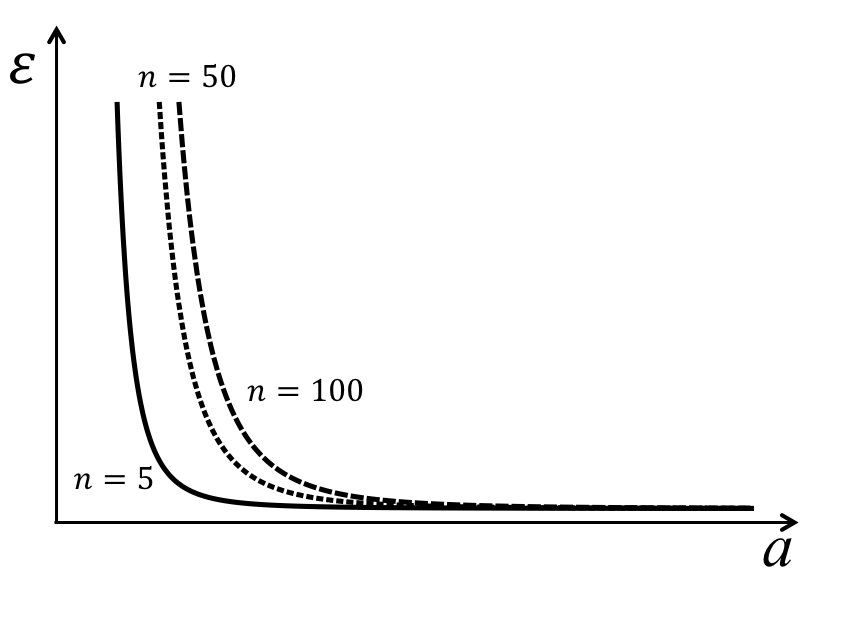}
\caption{Energy density of entanglement (see, Eqs. (\ref{Eent}-\ref{edensity})) for different values of the spatial mode of the scalar field.}
\label{figure14}
\end{figure}

\begin{figure}
\centering
\includegraphics[width=9cm]{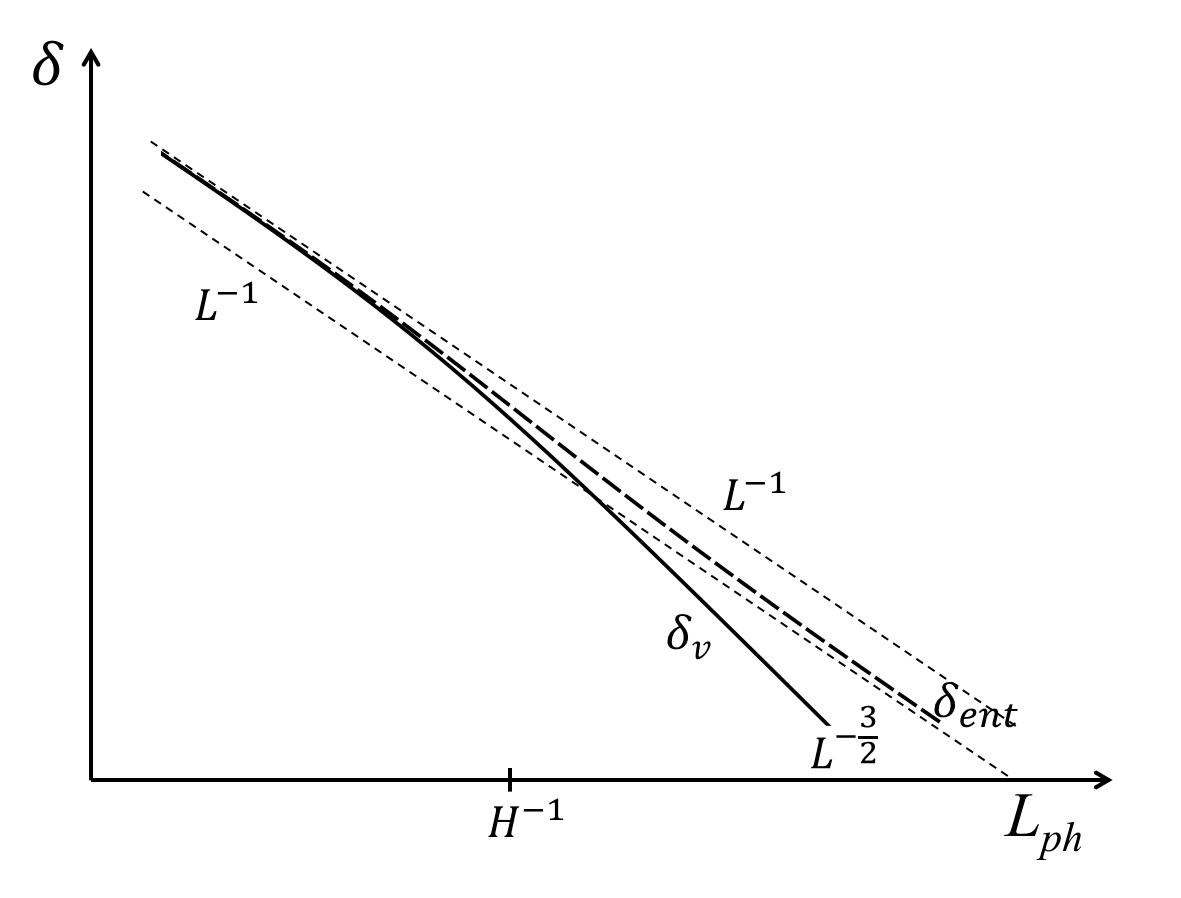}
\caption{Spectrum of quantum fluctuations for the vacuum state, $\delta_v$ given by (\ref{VSP01}) with, $L_\text{ph} = \frac{a}{n_\text{ph}}$ , and for the thermal state (\ref{RHO02}) derived from the entanglement between the two parent spacetimes where the scalar fields propagate.}
\label{figure15}
\end{figure}

Let us finally analyze the effects of the entanglement in the fluctuations of the scalar field, $\varphi$, which could shed some observable effect as well. Let us first consider the amplitude of the quantum fluctuations in the case where no entanglement is present. Then,
\be
\delta(n) = \frac{n^\frac{3}{2}}{2 \pi} \sqrt{\langle \hat{\varphi}_n^2 \rangle} ,
\ee
where,
\be
\langle \hat{\varphi}_n^2 \rangle = \text{Tr} \rho \hat{\varphi}^2_n = \langle 0 | \hat{\varphi}^2_n | 0 \rangle = \frac{1}{M \omega_n} ,
\ee
where it has been assumed that the field is in the vacuum state of the $b$, $b^\dag$ representation. Then, the fluctuations of the vacuum turns out to be
\be\label{VSP01}
\delta_\text{vac}(n) = \frac{n^\frac{3}{2} }{2 \pi a^\frac{3}{2} \left( 2 H^2 + \frac{n^2}{a^2}\right)^\frac{1}{4}} = \frac{n_\text{ph}^{\frac{3}{2}}} { 2 \pi (  2 H^2 + n_\text{ph}^{2} )^\frac{1}{4}} ,
\ee
where, $n_\text{ph} = \frac{n}{a}$. The spectrum (\ref{VSP01}) is like the spectrum of the vacuum fluctuations of Minkowski spacetime in terms however of the physical wave number $n_\text{ph}$, as it was expected due to the conformal invariance of the action of the spacetime with a conformally coupled massless scalar field. However, if the field is in the state (\ref{RHO02}) due to the entanglement between the universes, then, the spectrum of fluctuations would be given instead by
\be
\langle \hat{\varphi}^2_n \rangle = \frac{1}{M \omega_n} \left( 1 + N_n \right) ,
\ee
where $N_n$ is given in (\ref{N01}). Then,
\be
\frac{\delta_\text{ent}^2(n)}{\delta_\text{vac}^2(n)} = \frac{1}{1- e^{-\frac{1}{T_n}}} .
\ee
For large modes, $T_n \rightarrow 0$ and the spectrum of fluctuations coincide with the spectrum of the vacuum fluctuations, i.e. the effects of the entanglement between the universes are subdominant. However, for the lowest modes the departure from the spectrum of the vacuum fluctuations of a scalar field that propagates in an unentangled universe is significant. In the limit $n_\text{ph} \ll 1$, $N_n \approx T_n \sim \frac{H a}{n}$, and thus
\be
\frac{\delta_\text{ent}^2(n)}{\delta_\text{vac}^2(n)} \sim \frac{1}{n} e^{H t } .
\ee
It is also worth noticing that in terms of the physical length,
\beq
\delta_\text{ent} \sim \delta_\text{vac} \sim L_\text{ph}^{-1} ,  \ \ \ \ \ \ &L_\text{ph}& \ll H^{-1} , \\ \nonumber
\\
\delta_\text{ent} \sim \delta_\text{vac} \frac{H}{L_\text{ph}} \sim L_\text{ph}^{-1} ,  \ \ \ \ &L_\text{ph}& \gg H^{-1} .
\eeq
It means that the effect of the entanglement between the universes disappears for very large and very short distances. However, for physical distances of order of the Hubble length it becomes significant (see, Fig. \ref{figure15}). In that region the effect of the entanglement may have an important influence then in the power spectrum of the scalar field and thus it can leave distinguishable imprints in the  properties of the universes that can be, in principle, observable.

\section{Conclusions}

We have studied the cosmology of a homogeneous and isotropic spacetime endorsed with a conformally coupled massless scalar field. We have found six different solutions of the Friedmann equation that represent six different types of universes that are periodically distributed along the complex time axis. They are thus separated by Euclidean regions that represent quantum mechanical barriers. From a classical point of view one should just choose one of these solutions as the solution that represents the evolution of the universe.

Quantum mechanically, however, there is a non-zero probability for the universe to tunnel out through the quantum barrier and suffer a sudden transition to the state of another solution. We have computed the probability of tunneling in terms of the Euclidean action that corresponds to the Euclidean instanton that connects the two Lorentzian states of the universe. Moreover, two or more instantons can be matched to form double and generally speaking multiple instantons that would be the quantum origin of the entangled states between two or more universes.

We have obtained the quantum states of the branches of the universe, both in the formalism of the Wheeler-DeWitt equation and in the third quantization formalism. Within the former, we have analyzed the creation of a single universe from the spacetime foam, being this represented in the model by a quantum superposition of relative states between the quantum states of the spacetime and the quantum states of the scalar field. We have obtained that the quantum state of an evolved universe, for which the quantum fluctuations of the spacetime are frozen out, describes the quantum mechanics of a scalar field propagating in a  deSitter spacetime background. We have also shown that the superposition principle of the quantum mechanics of matter fields alone is an emergent feature of the semiclassical description of the universe.

We have shown that within the third quantization formalism the most natural way in which the universes are created is in entangled pair of branches, each one associated to the positive and negative values of the momenta conjugated to the scalar field. Their quantum states are given by the complex pairs of WKB solutions of the Wheeler-DeWitt equation. It parallels the creation of entangled pairs of particles with opposite momenta in a quantum field theory. However, the momenta conjugated to the scale factor is  related to the expansion rate of the universes and thus, the opposite signs of the momenta for the newborn universes correspond to the opposite expansion rates in terms of a common time variable $t$. Nevertheless, in terms of the WKB time variable, which would be the time variable measured by internal observer in their particle physics experiments,  the two universes are both contracting or expanding. It means that the universes are created in entangled pairs of both expanding or both contracting branches.

We have defined the appropriate vacuum state for the quantum description of the wave function of the multiverse. It corresponds to the stable \emph{no-universe} state for all values of the scale factor. However, in the invariant representation the multiverse is described in terms of pairs of interacting universes whose non-local interactions are high at the creation of the universe to decrease as the universes expand and become large parent universes like ours. The effective value of the  frequency of their quantum states, which is ultimately related to the Friedmann equation, is then significantly modified by the entanglement introducing a pre-inflationary stage in the evolution of the universes. It is worth noticing that a modification of the Friedmann equation that would entail the appearance of a pre-inflationary stage of the universe might produce a suppression of the lowest modes of the power spectrum of the CMB provided that inflation does not last for too long \cite{Bouhmadi2011, Scardigli2011}. It would be a way to test the effects of the multiverse in a more realistic model of the multiverse.

We have analyzed as well the effects of the entanglement between the modes of the scalar field that propagate in each spacetime of the entangled pair of universes. The quantum state of the scalar field in one of the universes turns out to be given by a quasi-thermal distribution whose time dependent temperature depends on the rate of entanglement between the two universes. The large modes of the scalar field are unaware of the entanglement between the two universes. For the lowest modes, however, the effect can be significant. It means that at short distances the effects of the inter-universal entanglement are negligible but they become important at distances of order of the Hubble length, which agrees with the result obtained in \cite{Mersini2008}.

We have computed the rate of particle creation. The largest modes of the scalar field remain in the vacuum state along the evolution of the field and there is thus no particle creation for these modes. For the lowest modes the number of created particles grows exponentially in time during the exponential expansion of the universe, a result that can be related to that obtained by Grishchuk and Sidorov \cite{Grishchuk1990} in the context of gravitational waves. We have computed the thermodynamical magnitudes that are associated to the thermal state of the scalar field that propagates in one of the universes. The density energy of entanglement becomes more important during the early stage of the evolution of the universe and it decreases to zero for a large parent universe like ours. Then, the effective value of the Friedmann equation would have three contributions. One is the initial energy density of the universe. The second is the energy density of entanglement between the spacetimes of the universes, and the third one is the energy of entanglement between the spatial modes of the original scalar field $\varphi(t)$ that propagates in the entangled spacetimes.

We have also computed the spectrum of fluctuations of the entangled scalar field and it has been compared with the spectrum of fluctuations of the vacuum state of a non-entangled scalar field. For the largest modes the effect would be unobservable. However, the inter-universal entanglement would have an important effect in the lower modes. In terms of the physical distance, the effect of entanglement is unobservable for distances both much larger and much shorter than the Hubble length. However, the spectrum of fluctuations of the field in the case it is entangled with the scalar field of the partner spacetime significantly departures from that of an unentangled universe at distances of order of the Hubble length, which would entail a distinguishable effect of the multiverse in the properties of the CMB.

This work opens the door for the search of these and other imprints of the multiverse in the properties of our universe by applying the same formalism to more realistic models of the universe and compare the outcomes with the astronomical data provided by the current and forthcoming space missions. Finally, let us also notice that the very existence of the multiverse is essentially derived from the subjacent physical theory, whether this is a string theory or a quantum theory of gravity. Therefore, if we finally find observable figures of the multiverse in the properties of our universe, then, they can be used to test these most fundamental theories.

\acknowledgments

\appendix

\section{Euclidean action integral (\ref{I01})}\label{A2}

Let us consider the integral
\beq\nn
I &=& \int_{a_-}^{a_+} \sqrt{- \bar{H}^2 a^4 + a^2 - \bar{C}} \\ \label{I1}
&=& \frac{1}{\bar{H}} \int_{a_-}^{a_+} \frac{\bar{H}^2 a^4 - a^2 + \bar{C}}{\sqrt{(a_+^2 - a )(a^2- a_-^2 )}} ,
\eeq
where $a_-^2 \equiv a_{1,-}$  and $a_+^2 \equiv a_{2,+}$ are given by Eqs. (\ref{a1minus}) and (\ref{a2plus}), respectively, and $\bar{H}$ and $\bar{C}$ are given by (\ref{Hbar}-\ref{Cbar}). Let us use the following change
\be
x^2 = 1 - \frac{a^2}{a_+^2} ,
\ee
then, the integral (\ref{I1}) transforms into
\be
I = \frac{1}{\bar{H} a_+ \sqrt{a_+^2 - a_-^2}} \int_0^\frac{1}{k} dx \frac{A_4 x^4 + A_2 x^2 + A_0 }{R} ,
\ee
where
\be
R = \sqrt{(1-x^2)(1-k^2 x^2)} ,
\ee
with, $k^2 = \frac{a_+^2}{a_+^2 - a_-^2}$, and 
\beq
A_4 &=& - \bar{H}^2 a_+^4 , \\
A_2 &=& 2 \bar{H}^2 a_+^4 - a_+^2 , \\ \label{A0eq}
A_0 &=& - \bar{H}^2 a_+^4 + a_+^2 - \bar{C}  .
\eeq
Taking into account that \cite{Baker1890}
\be\label{DER01}
0 = \int_0^\frac{1}{k} d(x R) = 3k^2 \int_0^\frac{1}{k} dx \frac{x^4}{R} - 2 (1+k^2) \int_0^\frac{1}{k} dx \frac{x^2}{R} + \int_0^\frac{1}{k} \frac{dx}{R} ,
\ee
we arrive at
\begin{widetext}
\be\label{I2B}
I = \frac{1}{k^2 \bar{H} a_+ \sqrt{a_+^2 - a_-^2}} \left\{  \left( \frac{2(1+k^2)}{3 k^2} A_4 + A_2 - \frac{A_4}{3}+ A_0 k^2\right)  F(q, k) - \left( \frac{2(1+k^2)}{3 k^2} A_4 + A_2\right) E(q,k) \right\} , 
\ee
\end{widetext}
where, $q = \arcsin\frac{1}{k}$. In the particular case for which $H_1 = H_2 \equiv H$ and $C_1 = C_2 \equiv C$, then (\ref{A0eq}) vanishes and
\be
I = -\frac{1}{3 k^2 (1-4 C H^2)^\frac{1}{4}} \left\{  C  F(\arcsin\frac{1}{k}, k) - a_+^2 E(\arcsin\frac{1}{k},k) \right\} .
\ee

\bibliographystyle{apsrev}
\bibliography{bibliography}

\end{document}